\documentclass[a4paper,11pt]{article}

\usepackage{jcappub}

\usepackage{cite}
\usepackage{amsmath,amssymb,epsfig}
\usepackage[dvipsnames]{xcolor}
\usepackage{mathrsfs}
\usepackage{color}
\usepackage{aas_macros}
\usepackage{ textcomp }
\usepackage{ wasysym }
\usepackage{fontawesome}
\usepackage{dirtytalk}

\newcommand{\beq}{\begin{equation}}
\newcommand{\eeq}{\end{equation}}

\renewcommand{\(}{\left(}
\renewcommand{\)}{\right)}
\renewcommand{\[}{\left[}
\renewcommand{\]}{\right]}
\renewcommand{\vec}[1]{\mathbf{#1}}
\newcommand{\vl}{\vec{l}}
\newcommand{\VL}{\vec{L}}

\newcommand{\vx}{\vec{x}}

\newcommand{\vtheta}{\boldsymbol{\theta}}

\newcommand{\vbeta}{\boldsymbol{\beta}}

\newcommand{\drm}{\mathrm{d}}
\newcommand{\ka}{\kappa}

\newcommand{\CMB}{\mathrm{CMB}}
\newcommand{\gal}{\mathrm{gal}}
\renewcommand{\max}{\mathrm{max}}
\newcommand{\mean}{\mathrm{mean}}
\newcommand{\lin}{\mathrm{lin}}
\newcommand{\li}{\mathrm{lim}}
\renewcommand{\d}{\mathrm{d}}

\title{Lensing corrections on galaxy-lensing cross correlations and galaxy-galaxy auto correlations}

\author[a,b]{Vanessa B\"ohm}
\author[a,b]{Chirag Modi}
\author[a,b]{Emanuele Castorina}

\affiliation{Berkeley Center for Cosmological Physics, University of California, Berkeley, CA 94720, USA}
\affiliation{Lawrence Berkeley National Laboratory, 1 Cyclotron Road, Berkeley, CA 93720, USA}

\emailAdd{vboehm@berkeley.edu}
\emailAdd{modichirag@berkeley.edu}
\emailAdd{ecastorina@berkeley.edu}

\abstract{We study the impact of lensing corrections on modeling cross correlations between CMB lensing and galaxies, cosmic shear and galaxies, and galaxies in different redshift bins. Estimating the importance of these corrections becomes necessary in the light of anticipated high-accuracy measurements of these observables. While higher order lensing corrections (sometimes also referred to as post Born corrections) have been shown to be negligibly small for lensing auto correlations, they have not been studied for cross correlations. We evaluate the contributing four-point functions without making use of the Limber approximation and compute line-of-sight integrals with the numerically stable and fast FFTlog formalism. We find that the relative size of lensing corrections depends on the respective redshift distributions of the lensing sources and galaxies, but that they are generally small for high signal-to-noise correlations. We point out that a full assessment and judgement of the importance of these corrections requires the inclusion of lensing Jacobian terms on the galaxy side. We identify these additional correction terms, but do not evaluate them due to their large number. We argue that they could be potentially important and suggest that their size should be measured in the future with ray-traced simulations. We make our code publicly available under \href{https://github.com/VMBoehm/lensing-corrections}{\faGithub}~\citep{code}.}
\begin{document}
\maketitle
\flushbottom

\section{Introduction}
\label{sec:intro}
Modeling of lensed observables relies on a number of assumptions. One of them, the Born approximation, states that it is sufficiently accurate to simply integrate lensing deflections along the line-of-sight instead of tracing them along the true photon geodesic. 
The list of works that have studied the validity of the Born approximation with analytical and numerical methods is long~\citep{2002CoorayHu,2006ShapiroCooray,2010A&AKrause,postBornPratten,postBornMarozzi,2015Calabrese,2017PhRvDPetri,2018JCAPFabbian} and has led to the consensus that it is extremely accurate for modeling shear and convergence power spectra. Higher order corrections to Born approximation  change, for example, the expected CMB lensing convergence power spectrum only at the sub-percent level\footnote{Terminology: what is commonly referred to as post Born corrections in the context of lensing observables is usually referred to as lensing corrections for other observables.}.
The success of the Born approximation is at first somewhat surprising because its validity is not due to the deflections being small - which they are not necessarily - but due to the fact that spatially uniform deflections have a large coherence scale. This leads to an efficient cancellation between otherwise large correction terms. 

Another intuitive phrasing for this finding was given by Ref.~\citep{kaiser_nonlinear_1995}: The applicability of the Born approximation is owed to the fact that it is used to model the statistics of the deflection field and that the statistical properties of the lenses along the perturbed path should not differ significantly from those along the unperturbed path. 

The relative smallness of lensing corrections is not guaranteed for all observables: Ref.~\citep{postBornPratten} found, for example, that post-Born corrections for the CMB lensing bispectrum can be of the order of the signal itself and Ref.~\citep{2017PhRvDPetri} pointed out the importance of higher order corrections for modeling the skewness and kurtosis of cosmic shear fields. Multiple deflection also source a curl component in the lensing deflection field, which could be detected with future experiments~\citep{2005PhRvDCooray,Namikawa_2012}. Recently, \citep{2019Fabbian} showed that lensing corrections are important for modeling cross-bispectra.

In this work, we want to study the validity of the Born approximation for cross correlations of lensing fields with other tracers of large scale structure. 
This is motivated by the recent vast literature showing that such cross correlations could be an extremely useful probe of neutrino masses, dark energy and other cosmological parameters \citep{2017Schmittfull,2018Yu,2018Giusarma}. 
Given the sub-\% measurements of cross-correlations that the next generation of cosmological surveys could achieve, it is therefore important to quantitatively check to what extent the Born approximation, assumed in all these forecasts, could provide an accurate descriptions of the data.

In the remainder of this paper we want to investigate the magnitude of Post-Born corrections to cross-correlations of lensing and galaxy fields, and how they depend on the relative redshifts of the observed galaxies. Our results will apply to lensing maps obtained from CMB data, galaxy shape catalogs or the intensity of the 21 cm line \citep{Foreman2018}.
We are also interested in relaxing the Limber approximation which assumes that only overlapping (in redshift) lensing planes are correlated, and we will indeed show that these often dropped terms are of the same magnitude as all the others.

We note that lensing is not the only higher-order correction to cross-correlations. A complete and consistent treatment of the problem up to fourth order in the linear density contrast, $\delta_\lin$, would require modeling the non-linear evolution of the matter density as well as the non-linear connection between the galaxy density and the dark matter density. While these corrections can be of the same order of magnitude or larger as the higher order lensing corrections considered here, a full treatment of all of these effects is beyond the scope of this paper.
Instead, we choose to focus on the question to what extent the cancellation found between higher-order lensing terms in the auto correlation can be recovered in cross correlations. By doing so we will use the Gaussian field $\delta_\lin$ as a small parameter and use a linear bias model to model the matter-halo connection. 

There has also been extensive work on higher order correction to the galaxy auto-correlation function~\citep{2014JCAPdiDio,2014PhRvDYoo,2014JCAPBertacca} and recently also line intensity mapping~\citep{2019JCAPJalivand}. Ref.~\citep{2014JCAPtSZcross} studied the impact of lensing corrections on the tSZ cross correlations and found them to be negligible.
While not the main scope of this paper, we also derive the expression for the auto power spectrum of galaxy number counts and evaluate it for different redshift distributions. 

This paper is organized as follows: we start by stating the leading order result and introducing notations in Section~\ref{sec:lead_order}. In Section~\ref{sec:formalism}, we give a schematic overview of the expected terms and cancellation. This is followed by expressions for higher order lensing corrections to both convergence and galaxy densities in Section~\ref{sec:higher_order}. In Section~\ref{sec:post_born} we start by listing all higher order terms that appear in the cross correlation and turn to their evaluation with and without Limber approximation in the subsequent subsections (Sections~\ref{sec:limber} and ~\ref{sec:no_limber}). In Section~\ref{sec:additional_terms}, we list all terms that involve the lensing Jacobian and point out their possible importance. In Section~\ref{sec:gal_gal} we turn to derive and evaluate corrections on the (auto) correlation between galaxy samples. We conclude in Section~\ref{sec:conclusions}. Details of the calculations can be found in the Appendices.
\section{Leading order expressions}
\label{sec:lead_order}
The leading order contribution to the observed two-dimensional galaxy density, $g(\vtheta)$, where $\vtheta$ is the angular position on the sky, is given by 
\begin{equation}
\label{eq:gal1}
g^{(1)}(\vtheta) = \int_0^{\chi_\max} \d \chi \[\frac{\d N}{\d z} \frac{\d z}{\d \chi}\] \delta_g(\vtheta,\chi)= \int_0^{\chi_\max} \d \chi\, W^{g}(\chi)\, \delta_g(\vtheta,\chi).
\end{equation}
We use the shorthand notation $W_g(\chi)$ for the normalized redshift-distribution of the galaxy sample
\begin{equation}
W_g(\chi) = \frac{\d N}{\d z} \frac{\d z}{\d \chi},
\end{equation}
where $\chi$ denotes the comoving distance.
By writing Eq.~\ref{eq:gal1} and all following expressions in terms of the three-dimensional galaxy density, $\delta_g$, we do not assume a specific bias model or order in the bias expansion.

In terms of lensing obervables, we will be working with the lensing convergence, $\kappa$, which is related to the lensing potential, $\phi$, through the Poisson equation,
\beq
\kappa(\vtheta) = - \frac{1}{2} \nabla^2 \phi(\vtheta).
\eeq
At leading order the lensing potential is modeled as a line-of-sight integral over the Weyl potential, $\Psi$,
\begin{equation}
\label{eq:lenspot}
\phi^{(1)}(\vtheta,\chi)= - 2 \int_0^{\chi} \d \chi'\, W_\kappa(\chi,\chi') \Psi(\vtheta,\chi'), 
\end{equation}
weighted by the lensing efficiency
\beq
\label{eq:Wkap}
W\[\chi,p(\chi')\]=  1/\chi \,  \int_{\chi}^{\chi_{\max}} \d \chi'\, p\(\chi'\) \frac{(\chi'-\chi)}{\chi'},
\eeq
with the distribution function of the sources $p(\chi)\d \chi=p(z) \d z$.
For a source plane at distance $\chi_s$, we have $p(\chi')=\delta_D(\chi'-\chi_s)$, and Eq.~\ref{eq:Wkap} simplifies to
\begin{equation}
W_\kappa(\chi,\chi_s)=\frac{\chi_s-\chi}{\chi \chi_s}\Theta(\chi_s-\chi),
\end{equation}
where $\Theta(\cdot)$ denotes the Heaviside step function. 

We further write the Poisson equation as
\begin{equation}
\nabla^2 \Psi(\vx,z) = \mathcal{A} (1+z) \delta_m(\vx,z),
\end{equation}
with the Weyl potential, $\Psi$, and the redshift independent prefactor defined as
\beq
\mathcal{A}=\frac{3}{2}\Omega_{m0} H_{0}^2.
\eeq
For notational brevity, we will make the $1/(1+z)$ in the Poisson equation part of the the lensing efficiency, i.e. modify Eq. \ref{eq:Wkap} to
\beq
W_\kappa(\chi,\chi_s) = \[1+z(\chi)\] \chi \frac{(\chi_s-\chi)}{\chi_s}\Theta(\chi_s-\chi).
\eeq
In the above equations and throughout this work we assume a flat Cosmology, i.e. $\Omega_K=0$ and delta function source redshifts.

On small scales, where the lensing signal is sourced by matter fluctuation that are small compared to the typical size of the lensing kernel, we can use the Poisson equation to express the lensing convergence in terms of the matter density contrast, $\delta_m$~\citep{2000ApJJain,2008PhRMunshi},
\begin{equation}
\label{eq:kap1}
\kappa^{(1)}(\vtheta)= \mathcal{A} \int_0^{\chi_s} \d \chi\, W_\kappa\[\chi,p(\chi_s)\] \delta_m(\vtheta,\chi).
\end{equation}
We are going to use this approximation in Section~\ref{sec:limber}, where we also apply the Limber approximation, because both approximations break down on similar scales. However, we are not going to use this approximation to compute our final results in Section~\ref{sec:no_limber}.
\section{General Formalism}
\label{sec:formalism}
To derive corrections on the leading order results, we will work with the traditional lensing expansion that assumes small deflections.
Since the aforementioned insensitivity to large coherent shifts is not reflected in this expansion, it appears in the calculation of two-point correlations as a cancellation between higher order terms. To illustrate this cancellation we schematically expand the convergence field $\kappa(\VL)$,
\begin{equation}
\label{eq:SimpleConvergenceExpansion}
\kappa\(\VL\) = \underbrace{\kappa^{(1)}(\VL)}_{\mathrm{Born}}+\underbrace{\kappa^{(2)}(\VL)}_{\mathcal{O}(d^2)}+\underbrace{\kappa^{(3)}(\VL)}_{\mathcal{O}(d^3)}+\mathcal{O}\(d^4\),
\end{equation}
where $\ka^{(1)}$ is the lensing convergence in the Born approximation and higher order corrections are labeled by their power in the lensing deflection $d=\nabla \phi$ (where we neglect curl modes). Note that we could replace $\kappa$ in Eq.~\ref{eq:SimpleConvergenceExpansion} by any other lensed field such as the projected galaxy density, $g$, emission line intensities, $\Delta$, or the CMB temperature and polarization fields, $T,E,B$. For these fields higher order terms would simply be referred to as lensing corrections. Post-Born corrections can be viewed as lensing corrections to the lensing field itself.

The 2-point correlator in harmonic space for convergence auto-spectrum is
\begin{align}
\label{eq:schem2ptauto}
\nonumber
\langle \kappa\(\VL\) \kappa\(\VL'\)\rangle= & \underbrace{\langle \kappa^{(1)}(\VL)\kappa^{(1)}(\VL')\rangle}_{\mathrm{Born}}+\underbrace{\langle \kappa^{(2)}(\VL)\kappa^{(1)}(\VL')\rangle}_{=0\ \mathrm{(Limber/Gaussian)}}+\underbrace{\langle \kappa^{(1)}(\VL)\kappa^{(2)}(\VL')\rangle}_{=0\ \mathrm{(Limber/Gaussian)}}\\
& + \underbrace{\langle \kappa^{(1)}(\VL)\kappa^{(3)}(\VL')\rangle+\langle \kappa^{(3)}(\VL)\kappa^{(1)}(\VL')\rangle+\langle \kappa^{(2)}(\VL)\kappa^{(2)}(\VL')\rangle}_{\mathrm{\approx 0}}+\mathcal{O}\(d^4\).
\end{align}
Third order terms in Eq.~\ref{eq:schem2ptauto} are absent for a Gaussian deflection field and vanish under the Limber approximation even for non-Gaussian fields (as we will show later). The smallness of the remaining terms relies on the cancellation of the rather large $(22)$ and $(13)+(31)$ contributions. This is analogous to, e.g., the cancellation of terms in modeling CMB lensing. Similar cancellations can also be found in standard perturbation theory for large-scale structure~\citep{1994JainBert,1996JainBert,scoccimarro96}.

Similarly, we can write the cross correlation between lensing convergence and projected galaxy density, $g$, as
\begin{align}
\label{eq:schem2ptcross}
\nonumber
\langle \kappa\(\VL\) g\(\VL'\)\rangle= & \underbrace{\langle \kappa^{(1)}(\VL)g^{(1)}(\VL')\rangle}_{\mathrm{Born}}+\underbrace{\langle \kappa^{(2)}(\VL)g^{(1)}(\VL')\rangle}_{=0\ \mathrm{(Limber/Gaussian)}}+\underbrace{\langle \kappa^{(1)}(\VL)g^{(2)}(\VL')\rangle}_{=0\ \mathrm{(Limber/Gaussian)}}\\
& + \underbrace{\langle \kappa^{(1)}(\VL)g^{(3)}(\VL')\rangle+\langle \kappa^{(3)}(\VL)g^{(1)}(\VL')\rangle+\langle \kappa^{(2)}(\VL)g^{(2)}(\VL')\rangle}_{\mathrm{\approx ?}}+\mathcal{O}\(d^4\),
\end{align}
and we see immediately that in order to recover the same number of terms as in the auto correlation, we have to take into account the lensing corrections to the galaxy field i.e. the fact that the galaxies themselves are observed at their lensed positions.\footnote{We will keep the same labeling of terms for the galaxy field as introduced for the lensing convergence, even though $g^{(1)}$ is zeroth order in the deflection, but motivated by the fact that $\kappa$ and $g$ are both of order $\nabla^2 \psi$, with $\psi$ being the gravitational potential.} This should not be confused with the effect of lensing on the galaxy shapes, known as cosmic shear.

\section{Higher order lensing  corrections}
\label{sec:higher_order}
Higher order lensing corrections to the convergence can be derived by perturbing the single lensing deflection $\Psi_{,a}(\vtheta,\chi)$ around the line-of sight-direction
\begin{align}
\nonumber
\Psi_{,a}\[\vtheta+\Delta \vtheta(\vtheta),\chi\] = &\Psi_{,a}(\vtheta,\chi) + \Psi_{,ab}(\vtheta,\chi) \Delta \vtheta_b(\vtheta,\chi)\\ & + \frac{1}{2}\Psi_{,abc}(\vtheta,\chi) \Delta \vtheta_b(\vtheta,\chi) \Delta \vtheta_c(\vtheta,\chi)+\mathcal{O}(\Delta \vtheta^3).
\end{align}
Here, we use $,a$ as short hand for $\nabla^a_{\vtheta}$, i.e., the $a$th component of the two dimensional angular derivative operator. Truncating at second order in the perturbation $\Delta \vtheta$ and using the expressions for $\Delta \vtheta^{(n)}$ that we derive in Appendix~\ref{app:lens_theory}, we get
\begin{align}
\nonumber
\label{eq:kappa-perturb}
\kappa(\vtheta) & &= \nabla^a_{\vtheta} \int^{\chi_\CMB}_{0} \d \chi\, W_\kappa(\chi,\chi_\CMB) & \left[  \Psi_{,a}(\vtheta \chi,\chi) + \Psi_{,ab}(\vtheta\chi,\chi)\phi_{,b}(\vtheta,\chi) \right. \\
\nonumber
&& &  +\frac{1}{2}\Psi_{,abc}(\vtheta\chi,\chi)\phi_{,b}(\vtheta,\chi)\phi_{,c}(\vtheta,\chi)\\
\nonumber
&&&   - 2\, \Psi_{,ab}(\vtheta\chi,\chi) \[\int_0^{\chi} \d \chi'  W_\kappa(\chi',\chi) \Psi_{,bc}(\vtheta \chi',\chi')\phi_{,c}(\vtheta,\chi')\]\\
&& &  \left.+\, \mathcal{O}(\Psi_{,a}^{4})\right]\\
&& = \kappa^{(1)}(\VL)+\kappa^{(2)}(\VL)+\kappa^{(3)}(\VL)&+\, \mathcal{O}(\Psi_{,a}^{4})
\end{align}
The intuitive interpretation of the above equation is that lensing changes not only the encountered lenses (Taylor expansion in $\Delta \vtheta^{(1)}$, sourcing the second and third term in the above equation), but also the subsequent path (fourth term in the above equation, proportional to $\Delta \vtheta^{(2)}$) and so on. Note that evaluating the derivative that we have moved outside of the integral will increase the number of terms. We will use small letters to number the terms at the same order, e.g.,
\beq
\kappa^{(3)}(\VL)=\kappa^{(3a)}(\VL)+\kappa^{(3b)}(\VL)
\eeq
or (after pulling the derivative inside)
\beq
\kappa^{(3)}(\VL)=\kappa^{(3A)}(\VL)+\kappa^{(3B)}(\VL)+\kappa^{(3C)}(\VL)+\kappa^{(3D)}(\VL).
\eeq
In analogy to Eq.~\ref{eq:kappa-perturb} lensing corrections to the observed galaxy field up to third order are
\begin{align}
\label{eq:gal-perturb}
\nonumber
g(\vtheta) = \int_0^{\chi_\max} & \d \chi\, W_g(\chi)  \left\{ J(\vtheta,\chi)  \(\delta_g(\vtheta\chi,\chi) + \delta_{g,a}(\vtheta\chi,\chi) \phi_{,a}(\vtheta,\chi) \right. \right. \\
& + \frac{1}{2} \delta_{g,ab}(\vtheta\chi,\chi) \phi_{,a}(\vtheta,\chi) \phi_{,b}(\vtheta,\chi)\\
& \left. \left. -2 \[\int_0^{\chi} \d \chi' W_\kappa(\chi',\chi) \phi_{,b}(\vtheta,\chi') \Psi_{,ab}(\vtheta\chi',\chi')\] \delta_{g,a}(\vtheta\chi,\chi) +1 \) -1 \right\} +\mathcal{O}(\delta_\lin^4).
\end{align}
Here, $J(\vtheta,\chi)$, denotes the Jacobian of the lens remapping (Eq.~\ref{simple_lensing})
\beq
J=\[(1-\kappa)^2-|\gamma|^2\]^{1-2.5s},
\eeq
where \say{$s$} is the change of number counts with magnitude at the magnitude limit of the survey. The effect of lensing magnification at lowest order is known as magnification bias~\citep{1980Turner,1984Turner,1988Webster,1988Fugmann,1989Narayan,1989Schneider,1995Villumsen}. For tracers without magnitude limit, like the CMB or the 21 cm radiation field, surface brightness conservation implies $J=1$.
The Jacobian itself contains higher order lensing corrections and needs to be expanded for our purposes,
\begin{equation}
J\equiv 1+\Delta J=1+J^{(1)}+J^{(2)}+J^{(3)}+\mathcal{O}(\Psi_{,a}^4).
\end{equation}
Detailed derivations and expressions for lensing corrections on number counts can be found in Appendix~\ref{app:gal_lens}.

Since $\Psi_{,aa} \simeq \delta_{\rm lin}$, as mentioned above, an expansion up to third order in the deflection field should also include second order and third order terms sourced by non-linear evolution~\citep{ForemanSenatore}(plus lensing corrections on these higher order terms), as well as a halo bias expansion. Accounting for all of these terms is beyond the scope of this work. Since our primary focus is to study the magnitude of the lensing corrections and possible internal cancellations, we will leave these terms aside, but note that they could potentially be of the same size or even bigger. Due to their different structure and physical source, an efficient cancellation between higher order terms sourced by different expansions is unlikely\footnote{However, Ref.~\citep{postBornPratten} identified such a cancellation in the CMB lensing bispectrum.}.

Another complication comes from the lensing Jacobian. It is non-linear in the lensing even without higher order lensing corrections. To simplify and organize the analysis, we ignore higher order terms in the lensing Jacobian in Section~\ref{sec:limber} and set $J=J^{(0)}=1$. In Section~\ref{sec:no_limber}, we take into account the magnification bias, sourced by correlating $J^{(1)}=5(s-0.4)\kappa$ with $\kappa^{(1)}$, because it is of the same order as the cross correlation itself. We discuss additional terms that arise in the presence of a lensing Jacobian in Section~\ref{sec:additional_terms}.
\section{Lensing corrections to galaxy-lensing cross correlations}
\label{sec:post_born}
With higher-order lensing corrections for both observables, galaxy density and lensing convergence, at hand we can now turn to computing the resulting corrections on their cross correlation. 
We split the assessment of these corrections in two sections. In the first section we make use of the Limber approximation, which states that unequal time/redshift correlations are negligible, i.e., that one can assume 
\beq
\langle A(z_1,k_1) B(z_2,k_2) \rangle \propto \delta_D(z_1-z_2)\delta_D(k_1-k_2),
\eeq
for two observables $A$ and $B$ at redshifts $z_1$ and $z_2$, respectively. We will also ignore the magnification bias for simplicity in this section and focus on examining the nature of the cancellations between higher order terms. After obtaining results in the Limber approximation, we will argue why this approximation is likely to break down for the expressions at hand. 
In the second section, we will compute and evaluate all correction terms without the Limber approximation and include lowest order magnification bias corrections.
To organize the results, we introduce the notation
\beq
C^{\ka g}(L) = C^{\ka g}_{11}(L)+C^{\ka g}_{12}(L)+C^{\ka g}_{21}(L)+C^{\ka g}_{22}(L)+C^{\ka g}_{31}(L)+C^{\ka g}_{13}(L)+\mathcal{O}(\delta_\lin^5),
\eeq
where subscripts label the order of each observable in $\delta_\lin$ (assuming $\Psi_{\,aa} \propto\delta_\lin$).
\subsection{Lensing corrections in the Limber approximation}
\label{sec:limber}

\subsubsection{Detailed expressions up to fourth order}
The leading order cross correlation between the convergence field and the galaxy field in Limber and flat sky approximation is given by
\beq
\label{CL1}
C^{\ka g}_{11}(L) = \mathcal{A} \int_0^{\chi_s} \d \chi \frac{W^{\kappa}(\chi,\chi_s) W_g(\chi)}{\chi^2} P_{mg}(L/\chi,\chi),
\eeq
where $\VL$ is the 2D harmonic wave vector on the sky and $L=|\VL|$ its modulus. 
Here, we have used the Poisson equation to relate the Weyl potential to the matter overdensity, assuming that derivatives of the potential along the line of sight integrate to zero. As mentioned in the beginning, this is a valid assumption when the derivative varies on scales much smaller then the typical size of the integration kernel. 

At next to leading order, the first contribution to the cross correlation signal comes from contracting the second order convergence term $\kappa^{(2)}$ with the leading order galaxy term $g^{(1)}$ (or vice versa). The resulting terms depend on the matter-halo cross bispectrum, $B_{mmg}$, which is zero if we ignore non-linear structure formation and the non-linear matter halo-connection. But even if one allows for a non-zero bispectrum, these terms still vanish under the Limber approximation:
for the bispectrum, the Limber approximation generalizes to
\begin{align}
\label{eq:bispeclimber}
\nonumber
\langle \delta_m(\VL_1, \chi_1)\delta_m(\VL_2, \chi_2)\delta_g(\VL_3, \chi_3)\rangle = & (2\pi)^2\delta_D(\VL_1+\VL_2+\VL_3)\frac{\delta_D(\chi_1-\chi_2)\delta_D(\chi_1-\chi_3)}{\chi_1^4} \times \\
& B_{mmg}(\VL_1/\chi_1, \VL_2/\chi_2, \VL_3/\chi_3; z(\chi_1)).
\end{align}
The delta functions in Eq.~\ref{eq:bispeclimber} collapse the lensing kernels in the expression for $\left<\kappa^{(2)}g^{(1)}\right>$
\beq
\delta_D(\chi_1{-}\chi_2)\delta_D(\chi_1{-}\chi_3) W_{\ka}(\chi_1,\chi_s)W_{\ka}(\chi_2,\chi_1)W_g(\chi_3) = W_{\ka}(\chi_1,\chi_s)\underbrace{W_{\ka}(\chi_1,\chi_1)}_{=0}W_g(\chi_1),
\eeq
and similarly for $\left< g^{(2)}\kappa^{(1)}\right>$.
Because of this twofold suppression, we will ignore these terms in the following.

At third order in the lensing convergence we get contributions from four terms,
\beq 
\langle \kappa^{(3)}(\VL) g^{(1)}(\VL')\rangle = \left\langle \[\kappa^{(3A)}(\VL)+\kappa^{(3B)}(\VL)+\kappa^{(3C)}(\VL)+\kappa^{(3D)}(\VL)\] g^{(1)}(\VL')\right\rangle
\eeq
while the lensing corrections to galaxy field results in two third order terms that we need to evaluate
\beq 
\langle \kappa^{(1)}(\VL) g^{(3)}(\VL')\rangle = \left\langle  \kappa^{(1)}(\VL) \[g^{(3a)}(\VL')+g^{(3b)}(\VL')\]\right\rangle.
\eeq
These expectation values involve the halo-matter four-point function, which consists of a Gaussian disconnected contribution and a connected trispectrum contribution. Sticking to the assumption of Gaussianity, we ignore the non-Gaussian trispectrum contribution\footnote{This should be a valid assumption at least for CMB lensing, which is sourced by lenses at relatively high redshifts, but might break down on small scales or for low source redshifts. The impact of bi- and trispectrum terms should be assessed with ray-traced simulations in the future.}. The Gaussian part can be decomposed under Limber into 
\begin{align} 
\nonumber
&\langle \delta_m(\VL_1,\chi_1)\delta_m(\VL_2,\chi_2)\delta_m(\VL_3,\chi_3) \delta_h(\VL_4,\chi_4)\rangle=\\
\nonumber
& (2\pi)^4 \delta_D(\VL_1+\VL_2)\delta_D(\VL_3+\VL_4)\frac{\delta_D(\chi_1-\chi_2)}{\chi_1^2}\frac{\delta_D(\chi_3-\chi_4)}{\chi_3^2} P_{mm}(L_1/\chi_1,\chi_1)P_{mg}(L_3/\chi_3,\chi_3)\\ &
+\(\VL_1 \leftrightarrow \VL_3, \chi_1 \leftrightarrow \chi_3\)+\(\VL_1 \leftrightarrow \VL_4, \chi_1 \leftrightarrow \chi_4\)
\end{align}
At first glance, this seems to result in 6x3 terms that contribute to the post-Born correction at this order. However, most of these terms vanish trivially in the Limber approximation because one of the lensing kernels becomes zero or because of the condition that multiple deflections can only be caused by lenses at different and ordered redshifts ($\chi_1>\chi_2>\chi_3$). Thus at third order in the lensing convergence we end up with two remaining terms, out of which one vanishes because of odd parity (see Appendix~\ref{app:add_terms_limber}) and the only remaining term is 
\begin{align}
\nonumber
\label{eq:Cl31}
C_{31D}^{(\kappa g)}(L)&=- 2 \ \mathcal{A}^3 \int_{\VL_1} \frac{\[\VL\cdot \VL_1\]^2}{L_1^4} & & \int_0^{\chi_\max} \drm \chi\, \frac{W_\kappa(\chi,\chi_s)W_g(\chi)}{\chi^2}\, P_{mg}(L/\chi,\chi) \\
& & &\int_0^\chi \drm \chi'\,  \frac{\[W_\kappa(\chi',\chi)\]^2}{\chi'^2} \, P_{mm}(L_1/\chi',\chi') \\
\nonumber
& = - \frac{\mathcal{A}^3 L^2}{2\pi} \int \d \ln  L_1 & & \int_0^{\chi_\max} \drm \chi\, \frac{W_\kappa(\chi,\chi_s)W_g(\chi)}{\chi^2}\, P_{mg}(L/\chi,\chi) \\
& & &\int_0^\chi \drm \chi'\,  \frac{\[W_\kappa(\chi',\chi)\]^2}{\chi'^2} \, P_{mm}(L_1/\chi',\chi').
\end{align}
Expanding the galaxy leg, we are again left with only one non-vanishing term
\begin{align}
\nonumber
C^{\kappa g}_{13b}(L') = & -2 \mathcal{A}^3 \int \frac{\d^2 \VL_1}{(2\pi)^2} \[\frac{\VL_1\cdot \VL'}{L_1^2}\]^2 \int_0^{\chi_s} \d \chi \frac{W_g(\chi)W_\kappa\(\chi,\chi_s\)}{\chi^2}\\
\label{Cl13}
& \int_0^{\chi} \d \chi' \frac{\[W_\kappa(\chi',\chi)\]^2}{\chi'^2} P_{mg}(L'/\chi,\chi)  P_{mm}(L_1/\chi',\chi').
\end{align}

Finally, we also need to correlate the two second order terms,
\begin{align}
\label{eq:Cl22}
\nonumber
C^{\kappa g}_{22}(l)= 4 \mathcal{A}^3 \int \frac{\d^2 \VL}{(2\pi)^2} \[\frac{\VL\cdot(\vl-\VL)}{|\VL-\vl|^2}\]^2 \frac{\VL\cdot \vl}{L^2} \int_0^{\chi_s} \d \chi \frac{W_g(\chi)W_\kappa\(\chi,\chi_s\)}{\chi^2} \\
\int_0^{\chi} \d \chi' \frac{\[W_\kappa(\chi',\chi)\]^2}{\chi'^2} P_{mh}(L/\chi,\chi)  P_{mm}(|\vl-\VL|/\chi',\chi').
\end{align}
At this point, non-linear corrections to the density and matter-halo connection could still be taken into account by using higher order expressions for the power spectra $P_{mm}, P_{mg}$ and $P_{gg}$, though this would not include mixed higher order terms and so for consistency, we refrain from doing that here.

To simplify the notation, we define the 2-dimensional matrix
\begin{align}
M(L,L')= & \int_0^{\chi_s} \d \chi \frac{W_g(\chi)W_\kappa\(\chi,\chi_s\)}{\chi^2} \int_0^{\chi} \d \chi' \frac{\[W_\kappa(\chi',\chi)\]^2}{\chi'^2} \frac{P_{mh}(L/\chi,\chi)}{L^2}  \frac{P_{mm}(L'/\chi',\chi')}{L'^4} \\\equiv &\int_0^{\chi_s} \d \chi  \int_0^{\chi} \d \chi'\, \mathcal{B}(\chi,\chi';\chi_s)\frac{P_{mh}(L/\chi,\chi)}{L^2}  \frac{P_{mm}(L'/\chi',\chi')}{L'^4}
\end{align}
which allows us to write the three non-zero terms in compact form
\begin{align}
C_{22}^{(\kappa g)}(L) & = 4 \mathcal{A}^3 \int \frac{\d^2 \VL'}{(2\pi)^2} \[\VL'\cdot\(\VL- \VL'\)\]^2 \[\VL\cdot\VL'\] M(L',|\VL-\VL'|)\\
C_{31}^{(\kappa g)}(L) + C_{13}^{(\kappa g)}(L) & = -4 \mathcal{A}^3 \int \frac{\d^2 \VL'}{(2\pi)^2} L^2 \[\VL\cdot\VL'\]^2 M(L,L').
\end{align}

\subsubsection{Understanding the cancellation}
Both terms in the above equation become very large, comparable or bigger than the signal at large $L$ (see Fig.~\ref{fig:CrossCorrectionLimber}).
In this form we can also easily see that these terms have the same structure as corrections to the auto-correlation (compare, e.g., Eqs.(33) and (36) in Ref.~\citep{2010A&AKrause}), which suggests a similar cancellation between them could happen.

In order to check this we first redefine $\VL'\rightarrow \VL-\VL'$,
\begin{equation}
C_{22}^{(\kappa g)}(L) = 4 \mathcal{A}^3 \int \frac{\d^2 \VL'}{(2\pi)^2} \[\VL'\cdot\(\VL- \VL'\)\]^2 \[\VL\cdot\(\VL- \VL'\)\] M(|\VL-\VL'|,L'),
\end{equation}
and then assume the power spectrum can be approximated by a power law, or a superposition thereof, over the relevant range of scales, $P_{mm}(k) \propto k^n$. In this limit the 31 term then reads
\begin{align}
    C_{31}^{(\kappa g)}(L) + C_{13}^{(\kappa g)}(L)  \simeq & -4 \mathcal{A}^3 \int_0^{\chi_s} \d \chi  \int_0^{\chi} \d \chi'\, \frac{\mathcal{B}(\chi,\chi';\chi_s)}{(\chi \chi')^n}  L^{2(n+1)}\int\frac{\d x \d\varphi}{(2 \pi)^2} x^{n-1}  \cos^2(\varphi)  \\ =& -4 \mathcal{A}^3 \int_0^{\chi_s} \d \chi  \int_0^{\chi} \d \chi'\, \frac{\mathcal{B}(\chi,\chi';\chi_s)}{(\chi \chi')^n}  L^{2(n+1)} \int\frac{\d x}{(4 \pi)} x^{n-1} 
\end{align}
where we have defined $x\equiv L'/L$ and $\varphi$ is the angle between $\VL'$ and $x$ axis. We see that for some choices of $n$ the lensing corrections receive large contribution when $x\rightarrow0$ or equivalently $L'\ll L$. As $L'$ becomes small, these modes become more and more a uniform shift applied to the modes of wavenumber $L$, and should therefore not be observable. 
The 22 term has a more complicated structure 
\begin{align}
   C_{22}^{(\kappa g)}(L)\simeq & 4 \mathcal{A}^3 \int_0^{\chi_s} \d \chi  \int_0^{\chi} \d \chi'\, \frac{\mathcal{B}(\chi,\chi';\chi_s)}{(\chi \chi')^n}  L^{2(n+1)} \notag \\ 
   & \int\frac{\d x \d\varphi}{(2 \pi)^2} x^{n+1}(1-\cos(\varphi)/x)^2(1-x \cos(\varphi))[1+x^2-2 x\cos(\varphi)]^{(n-2)/2}
\end{align}
which we then expand for small $x$ and integrate over $\varphi$
\begin{align}
     C_{22}^{(\kappa g)}(L)\simeq & 4 \mathcal{A}^3 \int_0^{\chi_s} \d \chi  \int_0^{\chi} \d \chi'\, \frac{\mathcal{B}(\chi,\chi';\chi_s)}{(\chi \chi')^n}  L^{2(n+1)} \int\frac{\d x}{(4 \pi)} [x^{n-1} +\mathcal{O}(x^{n+1})]\,.
\end{align}
The leading order large scale contributions indeed exactly cancels the one of the 31 piece and one is left with small subleading terms of $\mathcal{O}(x^{n+1})$.
Notice that the cancellations happens at the level of the integrand and that it is independent of the lensing and galaxy kernels.
Compared to the similar cancellation happening in the perturbative expansion for the density field \citep{scoccimarro96,jain_cosmological_1997,1996JainBert,Carrasco2013} the IR sensitivity of the individual terms is much stronger, and divergencies appear for $n\le 0$. 
While we have only shown that large scale shifts are unobservable for the first post Born correction and in Limber approximation, we expect this result to hold for the fully general case beyond Limber and in the fully non linear regime. 

\subsubsection{Numerical Evaluation}
To numerically evaluate the integrals we follow the trick used by Refs.~\citep{2010A&AKrause,Carrasco2013} and move the cancellation inside the integrand by rewriting,
\begin{align}
C_{22}^{(\kappa g)}(L) + C_{31}^{(\kappa g)}(L) = 4 \mathcal{A}^3 & \left[ \int \frac{\d^2 \VL'}{(2\pi)^2} \[\VL'\cdot\(\VL- \VL'\)\]^2 \[\VL\cdot\(\VL- \VL'\)\] \( M(|\VL-\VL'|,L') - M(L,L')\) \right. \\
& \left. + \int \frac{\d^2 \VL'}{(2\pi)^2} \(\[\VL'\cdot\(\VL- \VL'\)\]^2 \[\VL\cdot\(\VL- \VL'\)\]-L^2 \[\VL\cdot\VL'\]^2 \) M(L,L') \]\\
 = 4 \mathcal{A}^3 & \left[ \int \frac{\d^2 \VL'}{(2\pi)^2} \[\VL'\cdot\(\VL- \VL'\)\]^2 \[\VL\cdot\(\VL- \VL'\)\] \( M(|\VL-\VL'|,L') - M(L,L')\) \right. \\
& \left. + L^2 \int_0^{2\pi} \frac{\d \varphi }{(2\pi)^2} \(1+2\cos^2 \varphi\) \int \d L'  L'^5 M(L,L')\].
\end{align}
In the last line, we have assumed that $\VL$ is aligned with the external $x$-axis, i.e. $L=L_x$, and $\VL \cdot \VL'= L L' \cos \varphi$ with $\varphi$ being the angle between $\VL'$ and the $x$ coordinate axis. We further dropped terms that vanish under the angular integration.
\begin{figure}[ht!]
\includegraphics[width=0.49\textwidth]{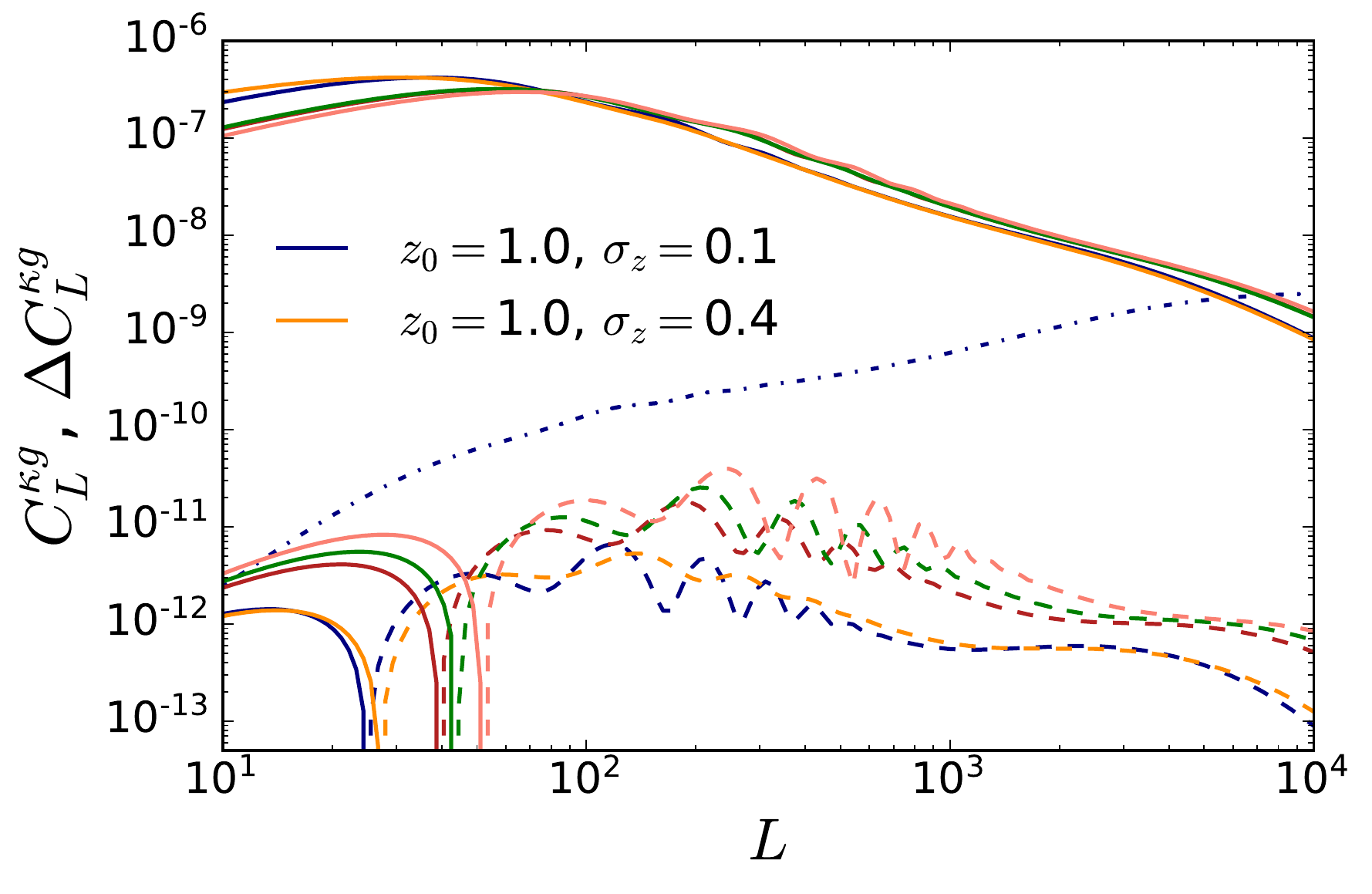}
\includegraphics[width=0.49\textwidth]{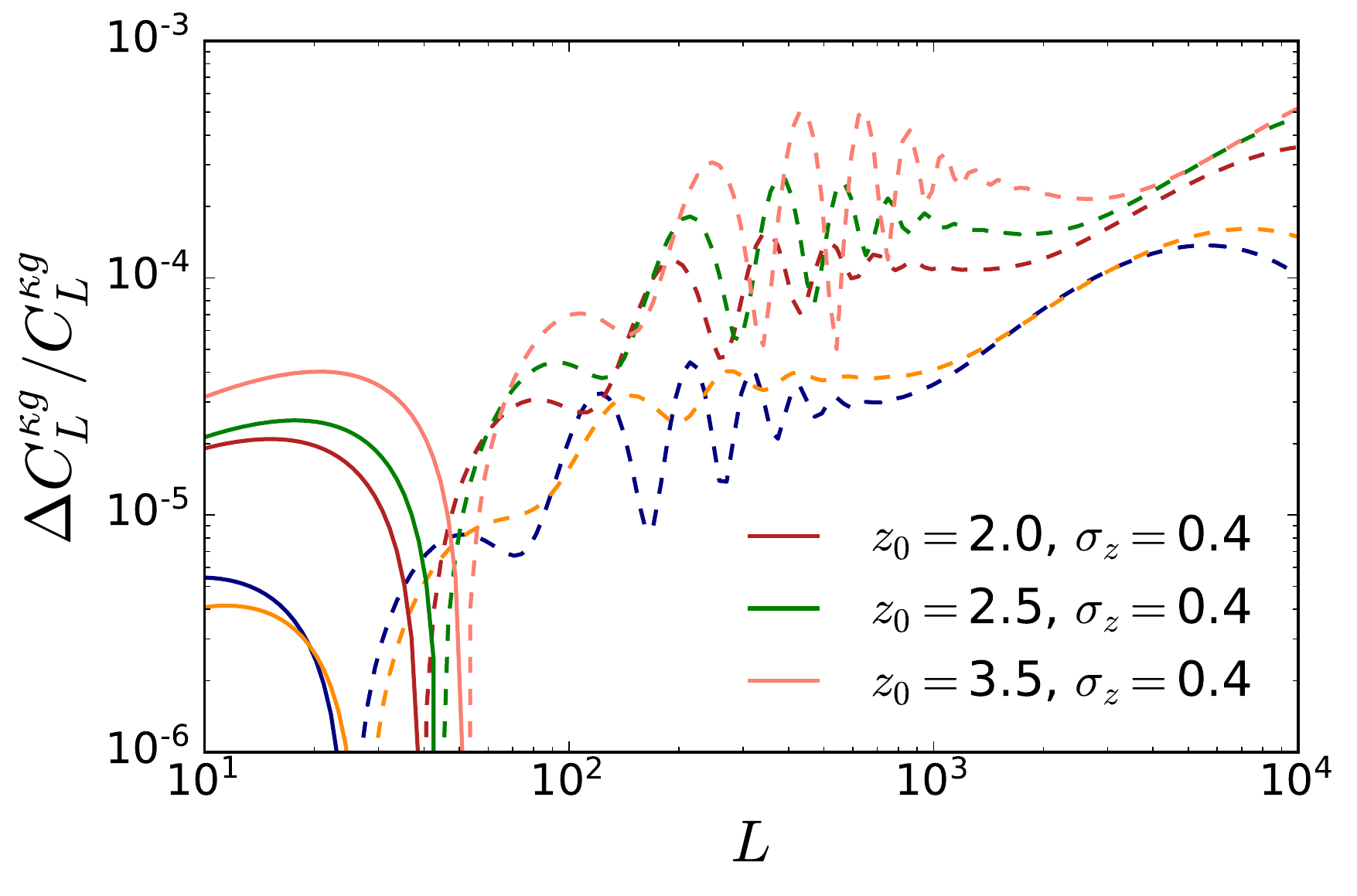}
\caption{\label{fig:CrossCorrectionLimber} Higher order lensing corrections to the galaxy- CMB lensing cross correlation for different Gaussian redshift distributions. On the left we plot the signals in solid lines, and the corrections in dashed (negative) and solid (positive) lines. To stress the importance of the cancellation, we plot the 31+13 terms in dashed-dotted. On the right we show the relative contribution of the correction to the signals, they are below 0.1\% up to $L=10 000$.}
\end{figure}

For the evaluation we adopt a $\Lambda$CDM cosmology with parameters $A_s=2.10732\cdot10^{-9}$, $h=0.68$, $k_{\mathrm{pivot}}=0.05$, $n_s=0.97$, $\omega_b=0.0225$, $\omega_{\mathrm{cdm}}=0.119$ and use a simple redshift dependent bias model of the form $b(z)=1+z$. We use the same cosmology and bias model throughout this paper. In Figure~\ref{fig:CrossCorrectionLimber} we plot the signal and correction terms for a correlation of the CMB lensing convergence with galaxy samples with Gaussian redshift distributions. We find a cancellation, similarly efficient as in the case of the lensing auto correlation.

\subsection{Lensing corrections without Limber approximation}
\label{sec:no_limber}
We now drop the Limber approximation and include magnification bias at lowest order in our calculations. 
Our motivations for dropping the Limber approximation are manifold:
\begin{enumerate}
\item The Limber approximation breaks down for projected scales that are similar in size to the projection kernel. The correction terms derived in the previous section integrate over varying scales and varying sizes of the kernel, such that this requirement is not always satisfied.
\item Other works, such as Ref.~\citep{2019JCAPJalivand}, have found that terms that vanish under Limber approximation can be of similar size or bigger than the sum of terms that are non-zero in Limber.
\item Depending on the relative redshifts of the observables, the Limber approximation can have a dramatic impact on scales $L<100$, e.g., causing a sign change of the signal (see e.g. Ref.~\citep{2017JCAPAssassi})
\end{enumerate}
We do not expect this full treatment to break the symmetries that lead to the cancellations that we analyzed in the previous section. 
However, the increased number of terms could add up to a significant correction, given the required precision for future cosmological measurements. 

In the following we define $C_l^{xy}$ as
\begin{equation}
    C_l^{xy}(\chi_{\max}, \chi'_{\max}) = \frac{2}{\pi} \int_0^{\chi_\max} \d \chi\, \int_0^{\chi'_\max} \d \chi'  W_x(\chi) W_y(\chi') \int \mathrm{dln}{k}\, j_l(k\chi) j_l(k \chi')  \[k^3 P_{xy}(k)\],
\end{equation}
where $j_l$ are spherical Bessel functions and kernels $W_x$ for different fields $x/y$ are listed in Table~\ref{tab:table1}. For the power spectrum, we use the linear 3D power spectrum of density fluctuations at $z=0$, $P_{\delta_\lin}$, and multiply by $\mathcal{A} k^2$ where necessary to convert $\delta \rightarrow \Psi$.
\begin{table}[ht!]
  \begin{center}
    \begin{tabular}{l|l}
      $x$ & $W_x(\chi)$ \\
      \hline
      $\delta_g$ & $D_+(\chi)b(\chi) \delta_D(\chi_\max)$ \\
      $g$      & $W_g(\chi) b(\chi) D_+(\chi)$\\
      $\Psi$   & $D_+(\chi)\delta_D(\chi_\max) $ \\
      $\phi$   & $D_+(\chi)W_\kappa(\chi,\chi_{\max}) $\\
    \end{tabular}
    \caption{\label{tab:table1} Kernels used in the $C_l$ computation in this section. $D_+$ denotes the growth function of linear perturbations, $b(\chi)$ is a redshift dependent bias (here $b=1+z$) and $\delta_K$ is the Dirac delta function. The lensing and redshift kernels ($W_g$, $W_\kappa(\chi,\chi_s)$) were defined in Section~\ref{sec:lead_order}.}
  \end{center}
\end{table}
Evaluating these line-of-sight integrals can be computationally challenging and expensive, due to the oscillatory behaviour of the Bessel functions. We use the recently proposed FFTlog formalism~\citep{2018JCAPSimonovic,2017JCAPAssassi} to evaluate our post-Limber expressions, because it is both fast and numerically stable. We test our implementation by reproducing correlation functions for CMB lensing and number counts produced with state of the art Boltzmann solvers~\citep{CAMB,CLASS}. We also independently double coded all expressions and debugged by cross checking the results.

\subsubsection{Detailed expression up to fourth order}
We start by stating the leading order results. Including magnification bias, we get two terms, the cross correlation between the galaxy field and lensing convergence 
\begin{equation} 
\label{eq:lo}
C^{\kappa g}_L= - L^2 \int_0^{\chi_s} \d \chi\, W_\kappa(\chi,\chi_s) \int_0^{\chi_\max} \d \chi'\,W_g(\chi')\, C_L^{\Psi \delta_g} (\chi,\chi')
\end{equation}
and the correlation between the first order lensing Jacobian ($J^{(1)}=-2 \int \d \chi W_g(\chi)\kappa(\chi)$) and the lensing convergence
\begin{equation}
\label{eq:magbias}
C_{1J1}^{\kappa g}(L)=  5(s-0.4)\, L^4 \int_0^{\chi_\max} \d \chi\, W_g(\chi) \int_0^{\chi_s} \d \chi'\, W_\kappa(\chi',\chi_s) \int_0^{\chi} \d \chi''\, W_\kappa(\chi'',\chi) C_L^{\Psi\Psi}(\chi',\chi'').
\end{equation}
For the lensing correction terms, we only consider terms at lowest order in the lensing Jacobian $(J^{(0)}=1)$.
These are the same correlations as in the previous section,
\begin{align}
\nonumber
C_{22}^{(\kappa g)}(L) = - & \int \frac{\d^2 \vl}{(2 \pi)^2} \[\VL\cdot \vl\] \[\vl\cdot\(\VL-\vl\)\]^2  \int_0^{\chi_s} \d \chi \int_0^{\chi_\max} \d \chi' \, W_g(\chi') W_\kappa(\chi, \chi_s) \\
\label{eq:22A}
&\[C_l^{\delta_g \Psi}(\chi',\chi) C^{\phi\phi}_{|\VL-\vl|}(\chi',\chi)\right. \\
\label{eq:22B}
& \left. +\, C_l^{\phi \Psi}(\chi',\chi) C^{\delta_g\phi}_{|\VL-\vl|}(\chi',\chi)\]
\end{align}
 Imposing Limber by requiring $\chi=\chi'$, sets the second term to zero and we recover our former result Eq.~\ref{eq:Cl22}. The second term vanishes because the Limber approximation sets source ($\phi$) and lens ($\psi$) to the same redshift, but the upto $\chi$ projected lensing field $\phi(\chi)$ gets no contribution from $\Psi(\chi)$, therefore $C_l^{\phi \Psi}(\chi,\chi)=0$.

When correlating the first order galaxy term with the first third order lensing term, we get
\begin{align}
\label{eq:31aA}
C_{31a}^{(\kappa g)}(L) &= \frac{1}{2} L^2 \int \frac{\d^2 \vl}{(2 \pi)^2} \[\VL\cdot \vl\]^2 \int_0^{\chi_s} \d \chi \int_0^{\chi_\max} \d \chi' \, W_g(\chi') W_\kappa(\chi, \chi_s) 
 C_L^{\delta_g \Psi}(\chi',\chi) C^{\phi\phi}_{l}(\chi)\\
 \label{eq:31aB}
 &  + \int \frac{\d^2 \vl}{(2 \pi)^2} \[\VL\cdot \vl\]^2 l^2 \int_0^{\chi_s} \d \chi \int_0^{\chi_\max} \d \chi' \, W_g(\chi') W_\kappa(\chi, \chi_s) 
 C_l^{\Psi \phi}(\chi) C^{\phi\delta_g}_{L}(\chi,\chi').
\end{align}
Similarly, contracting the first order lensing term with the first third order galaxy term, results in
\begin{align}
\label{eq:13a}
C_{13a}^{(\kappa g)}(L) = \frac{1}{2} L^2 \int \frac{\d^2 \vl}{(2 \pi)^2} \[\VL\cdot \vl\]^2 \int_0^{\chi_s} \d \chi \int_0^{\chi_\max} \d \chi' \, W_g(\chi') W_\kappa(\chi, \chi_s) 
 C_L^{\delta_h \Psi}(\chi',\chi) C^{\phi\phi}_{l}(\chi').
\end{align}
The sum of these two terms is the same as Eq.~\ref{eq:Cl31} after imposing the approximation.

Finally, correlating the second third order lensing term with the first order galaxy term, gives
\begin{align}
\nonumber
\label{eq:31b}
C_{31b}^{(\kappa g)}(L) = & 2 \int \frac{\d^2 \vl}{(2 \pi)^2} \[\VL\cdot \vl\]^2 l^2 \int_0^{\chi_s} \d \chi\, W_\kappa(\chi, \chi_s) \int_0^{\chi_\max} \d \chi' \, W_g(\chi') \\
&
\int_0^\chi   \d \chi''\, W_\kappa(\chi'',\chi) C_l^{\Psi \Psi}(\chi'',\chi) C^{\phi\delta_g}_{L}(\chi'',\chi')
\end{align}
and 
\begin{align}
\nonumber
C_{13b}^{(\kappa g)}(L) =& -2 \int \frac{\d^2 \vl}{(2 \pi)^2} \[\VL\cdot \vl\]^2 L^2 \int_0^{\chi_s} \d \chi\, W_\kappa(\chi, \chi_s) \int_0^{\chi_\max} \d \chi' \, W_g(\chi') \\
\label{eq:13b}
&
\int_0^{\chi'}   \d \chi''\, W_\kappa(\chi'',\chi') C_L^{\Psi \Psi}(\chi'',\chi) C^{\phi\delta_h}_{l}(\chi'',\chi'),
\end{align}
which are both zero in Limber because $W_\kappa(\chi,\chi)=0$.

\subsubsection{Results}
\begin{figure}[ht!]
\begin{center}
\includegraphics[width=0.32\textwidth]{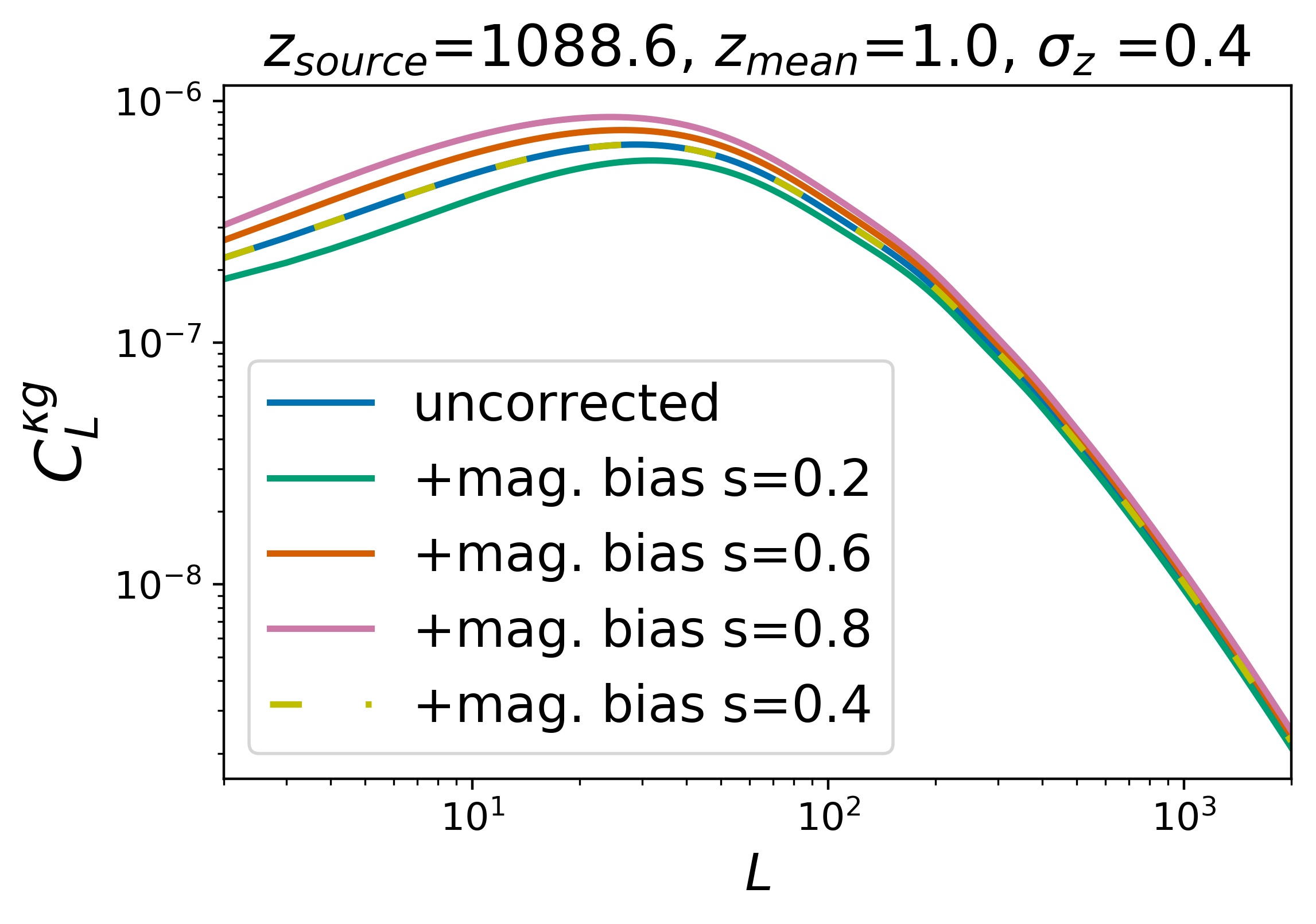}
\includegraphics[width=0.32\textwidth]{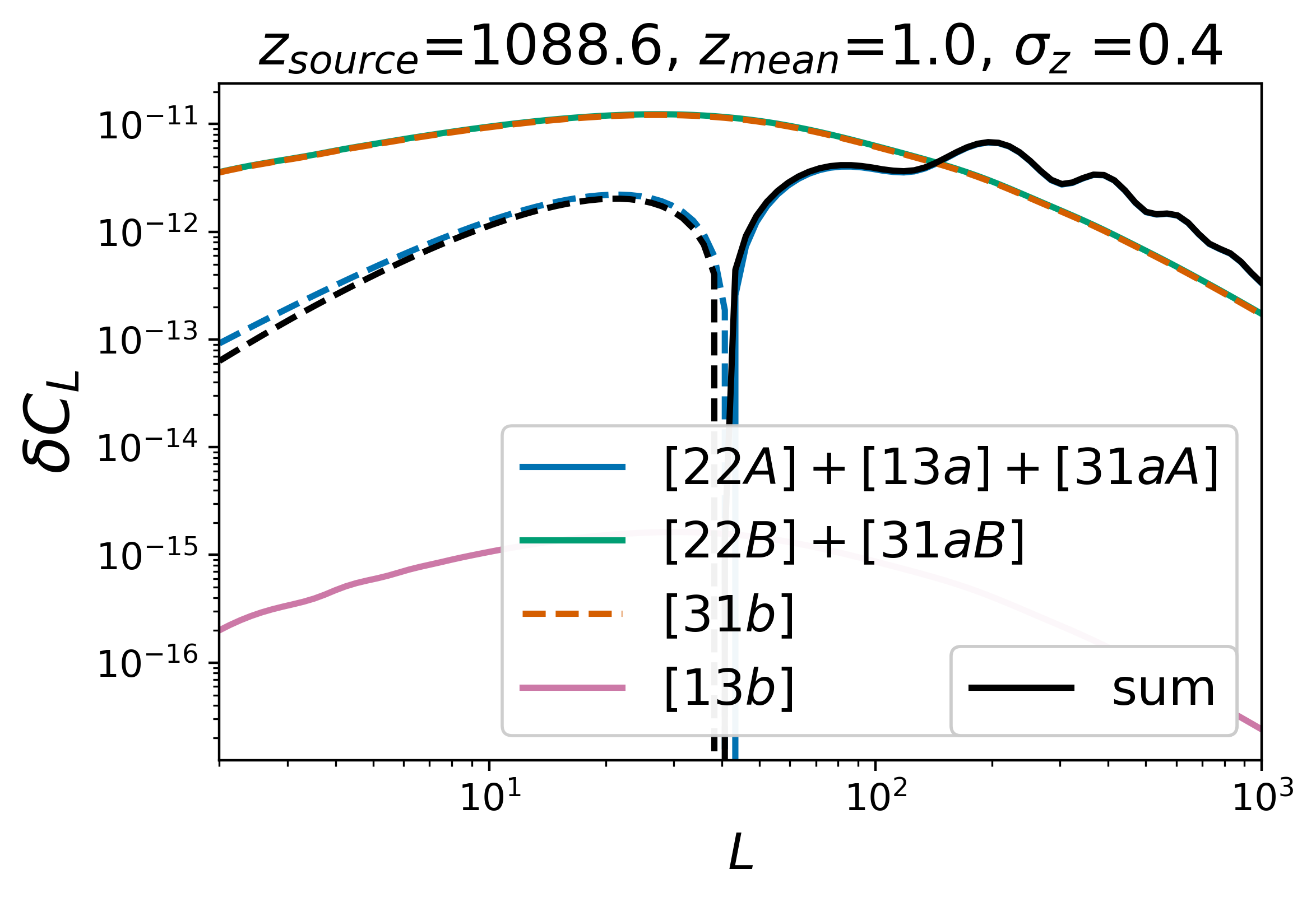}
\includegraphics[width=0.32\textwidth]{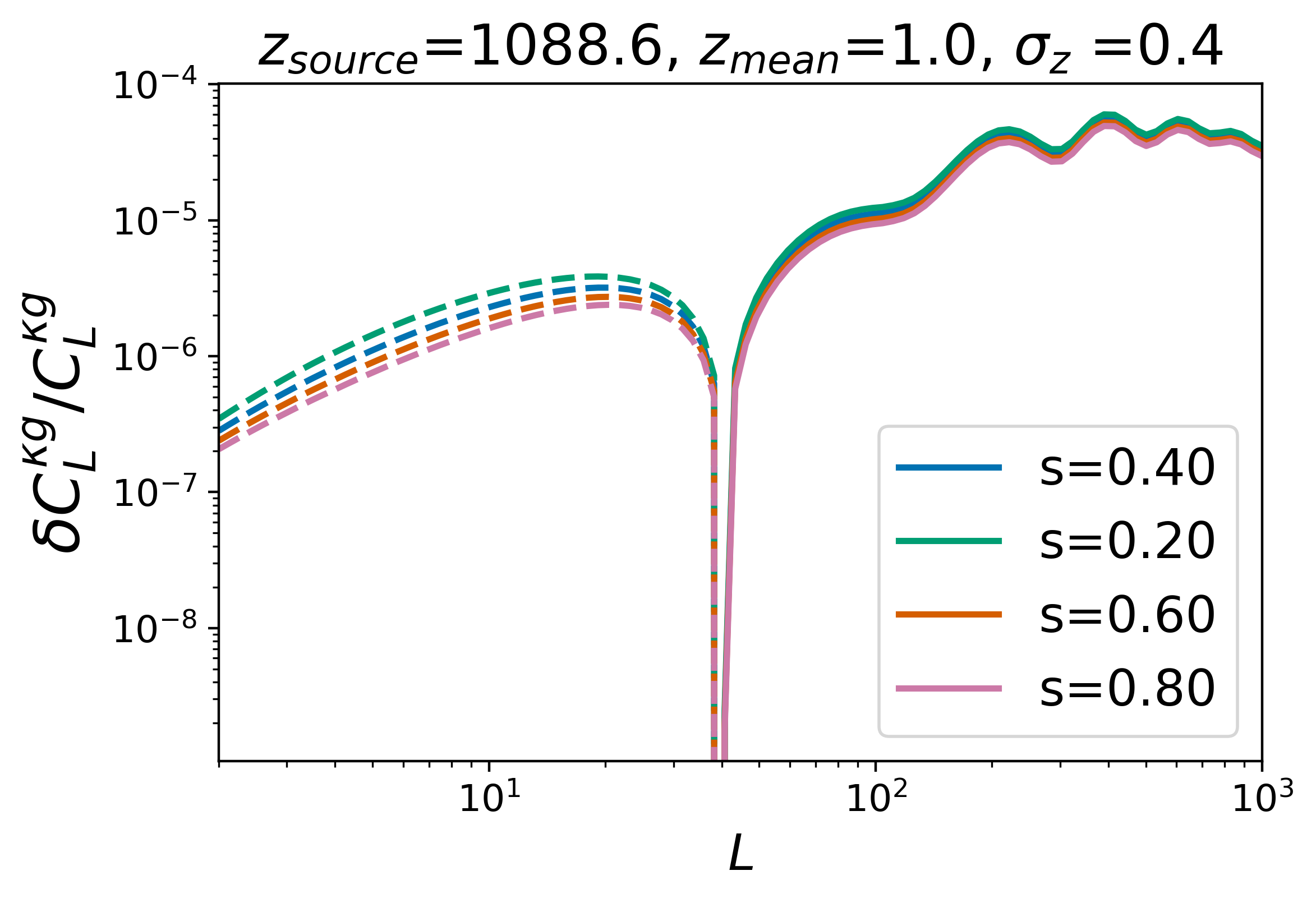}
\includegraphics[width=0.32\textwidth]{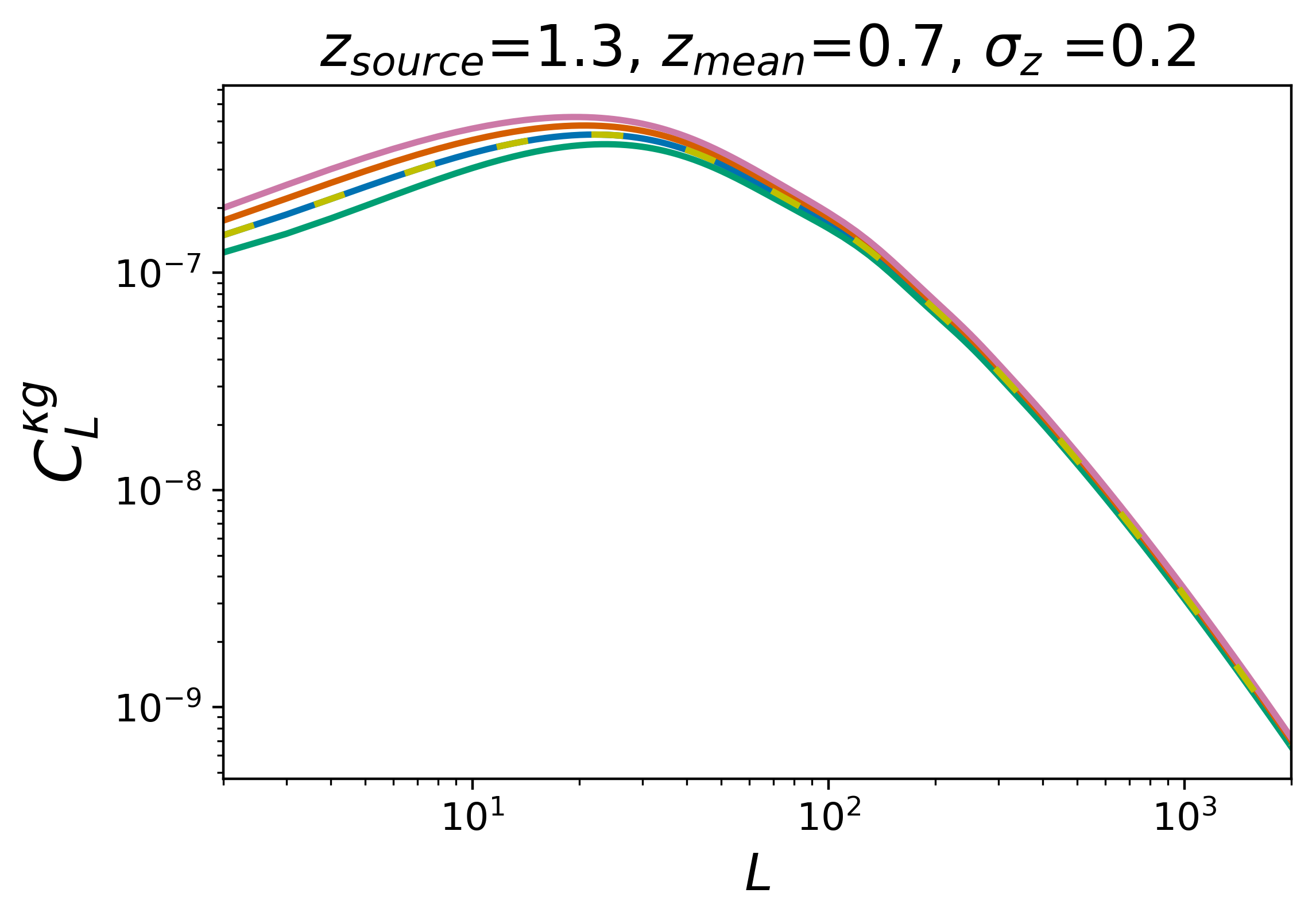}
\includegraphics[width=0.32\textwidth]{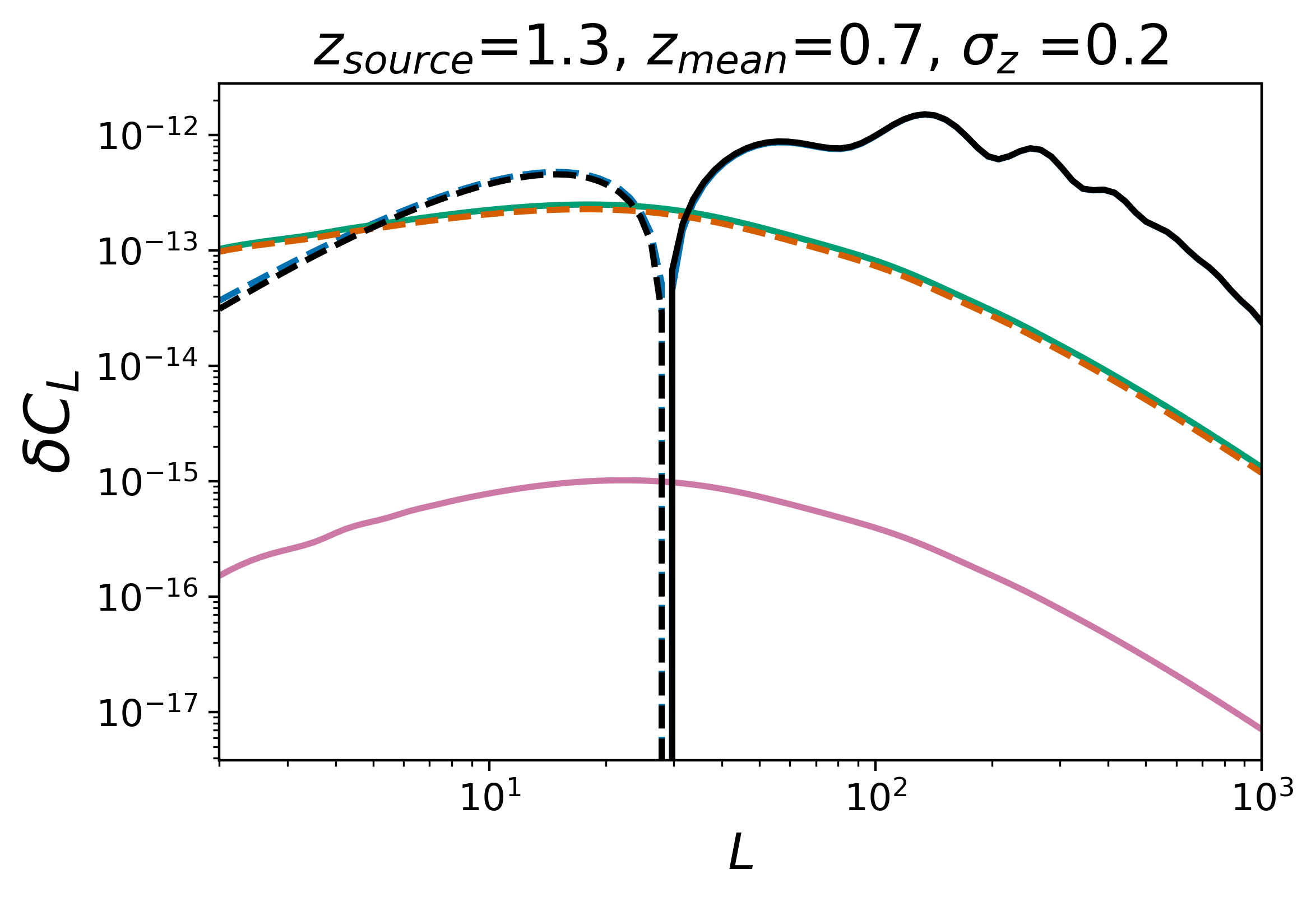}
\includegraphics[width=0.32\textwidth]{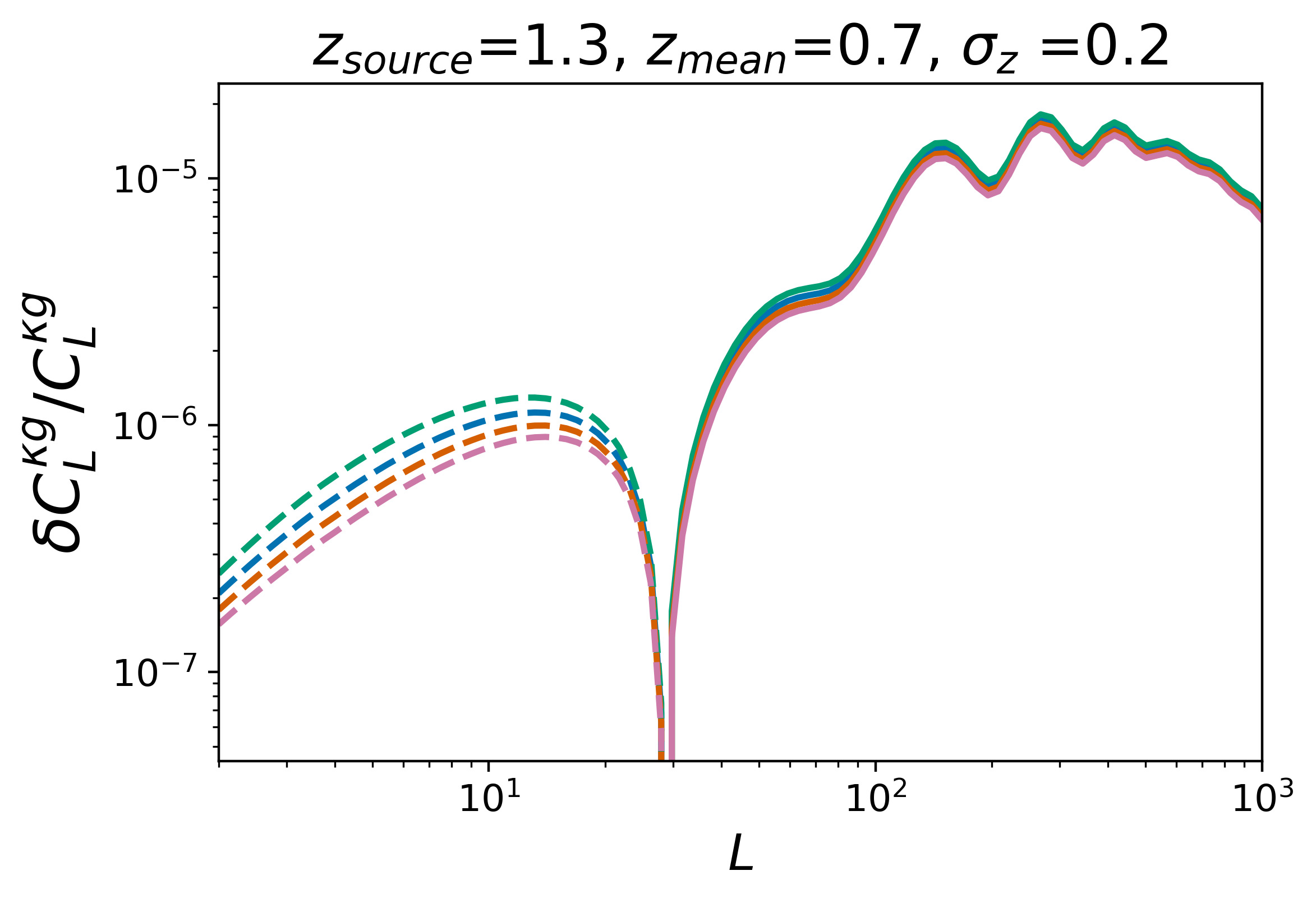}
\includegraphics[width=0.32\textwidth]{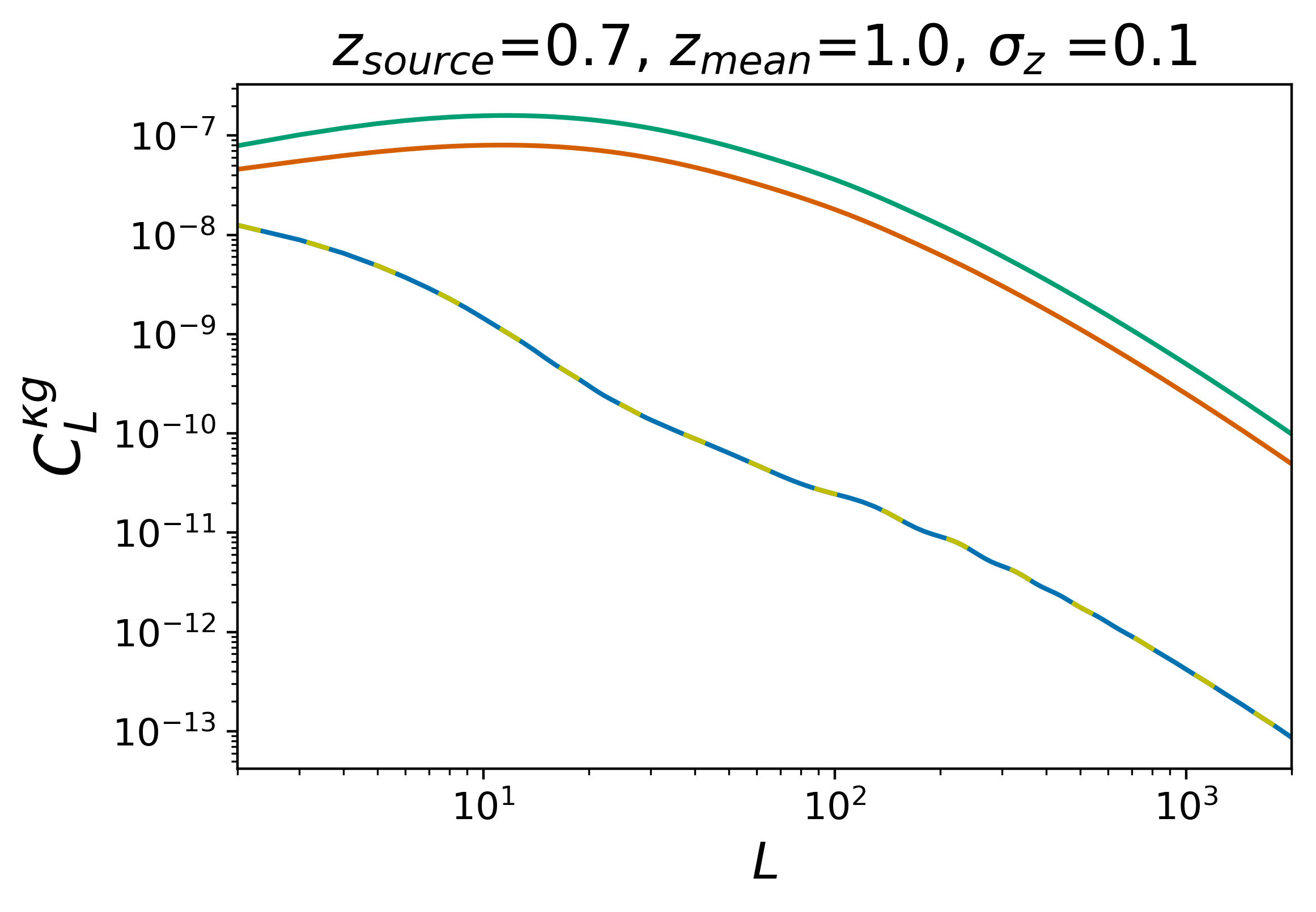}
\includegraphics[width=0.32\textwidth]{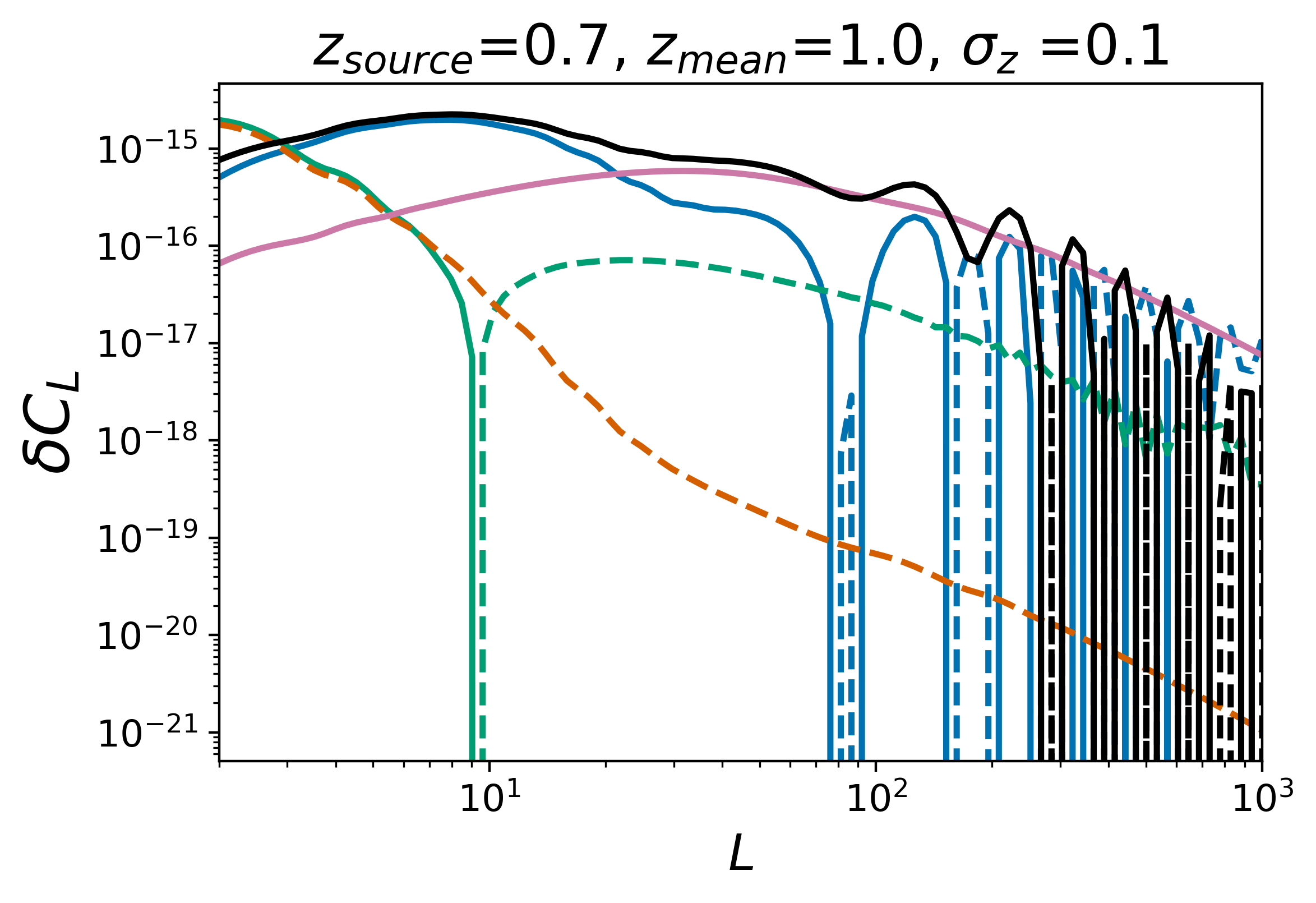}
\includegraphics[width=0.32\textwidth]{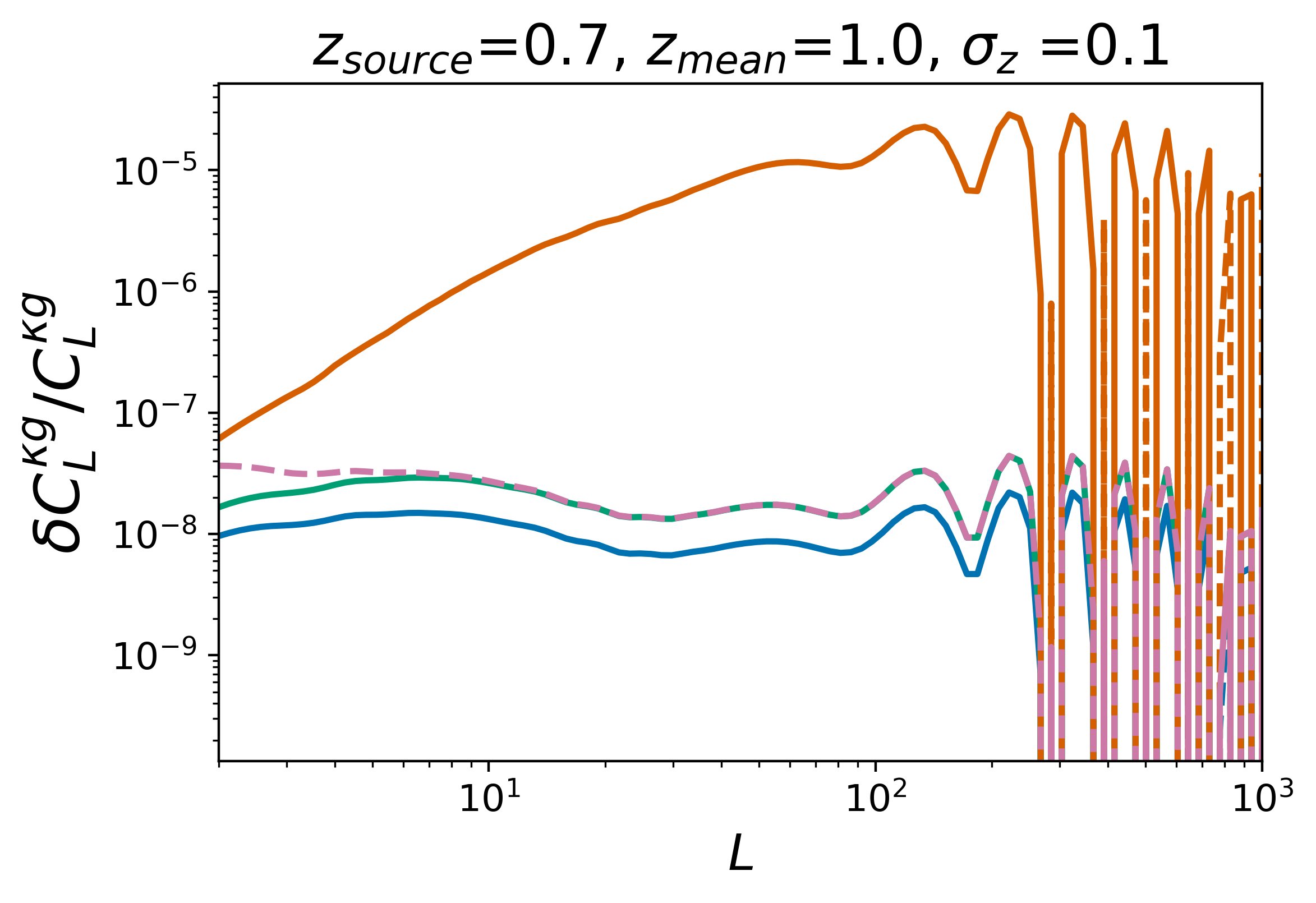}
\end{center}
\caption{\label{fig:CrossCorrectionnoLimber} Leading order results (Eq.\ref{eq:lo}+Eq.\ref{eq:magbias}) for different slope parameters $s$ (left column) and corresponding higher order lensing corrections (Eqs.\ref{eq:22A}-\ref{eq:13b}) (middle column) for three different redshift configurations. The leading order cross correlations lie several orders of magnitude above the lensing corrections for all source and galaxy distributions considered (right column). Our highest assumed value of $s=0.8$ is unlikely to come from the flux limit alone, but can arise in the presence of an additional size bias~\citep{size_bias}.}
\end{figure}

We evaluate the above expressions for different source redshifts and redshift kernels.
In all examples we assume for simplicity Gaussian galaxy distributions characterized by a central redshift and a variance. We choose three different settings: Correlating CMB lensing with galaxies of central redshift $z_\mean=1$ and width $\sigma_z =0.4$ (the same setting was used in the evaluation under Limber approximation in Section~\ref{sec:limber}), correlating galaxy lensing with source redshift $z_s=1.3$ with a galaxy distribution peaking at $z_\mean=0.7$ and $\sigma_z=0.2$ (achieving a large overlap of galaxy and lensing kernel) and finally a setting in which the signal is dominated by magnification bias, i.e., where the source redshift lies in front of the galaxy distribution.
\begin{figure}[ht!]
\begin{center}
\includegraphics[width=0.32\textwidth]{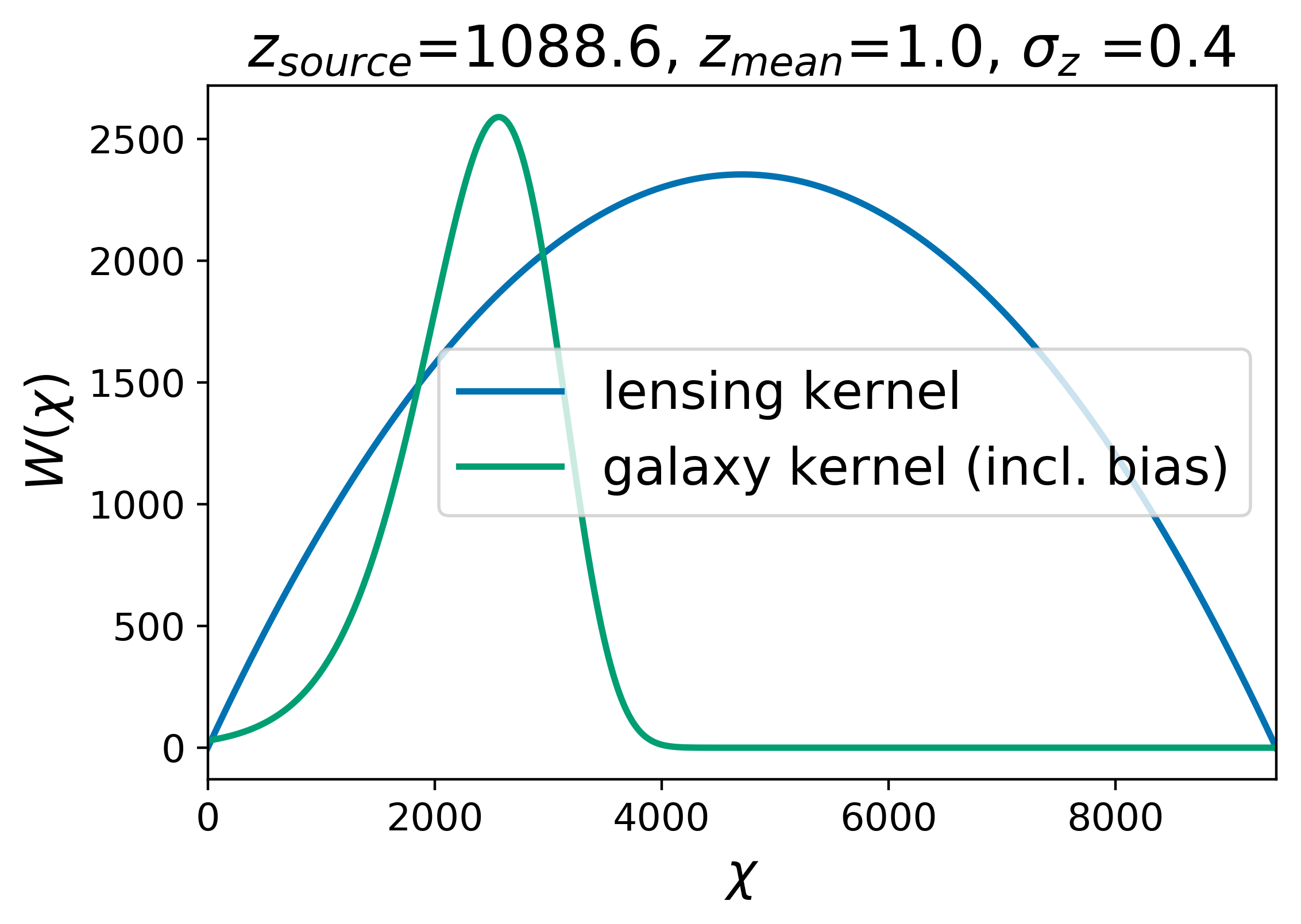}
\includegraphics[width=0.32\textwidth]{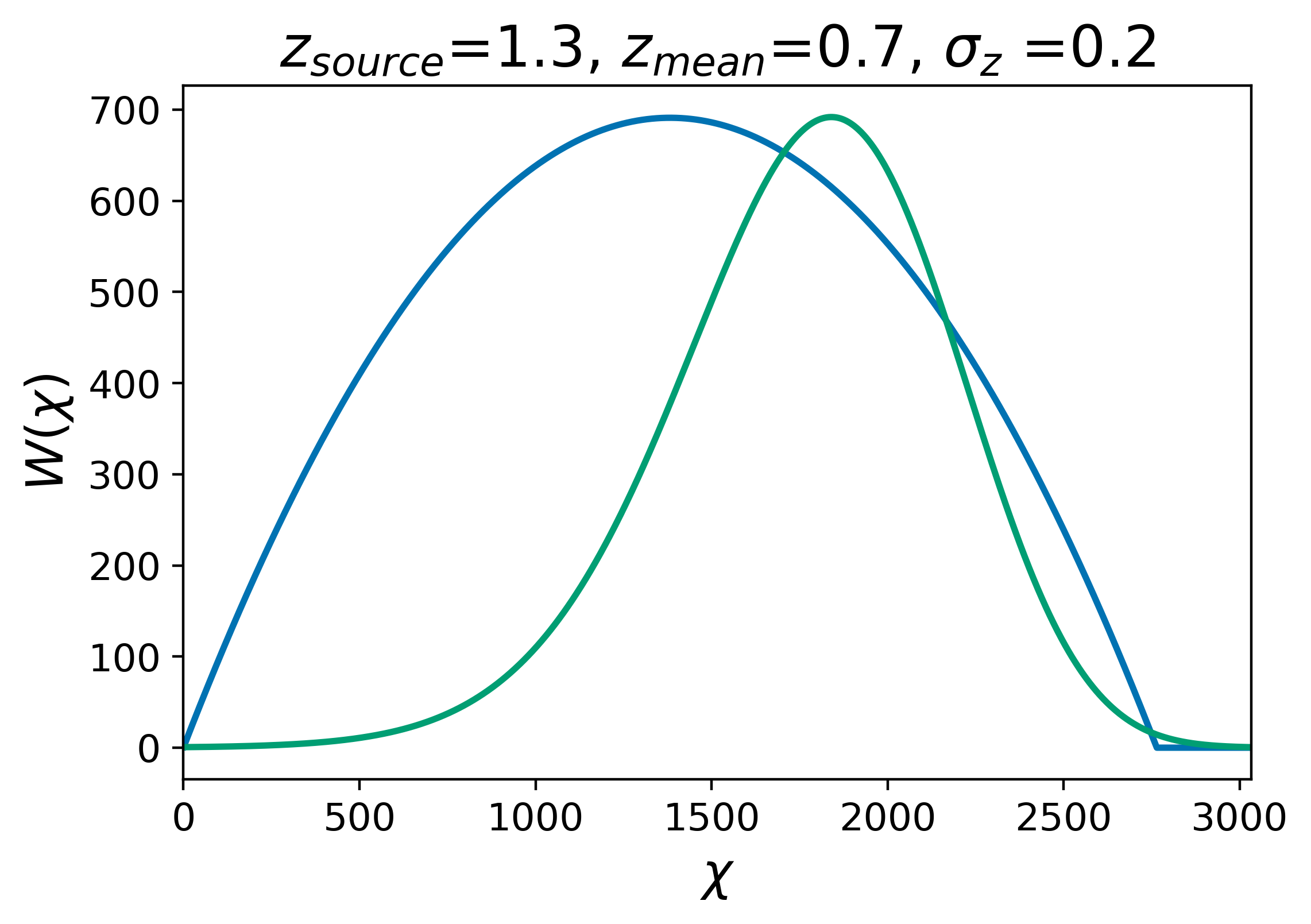}
\includegraphics[width=0.32\textwidth]{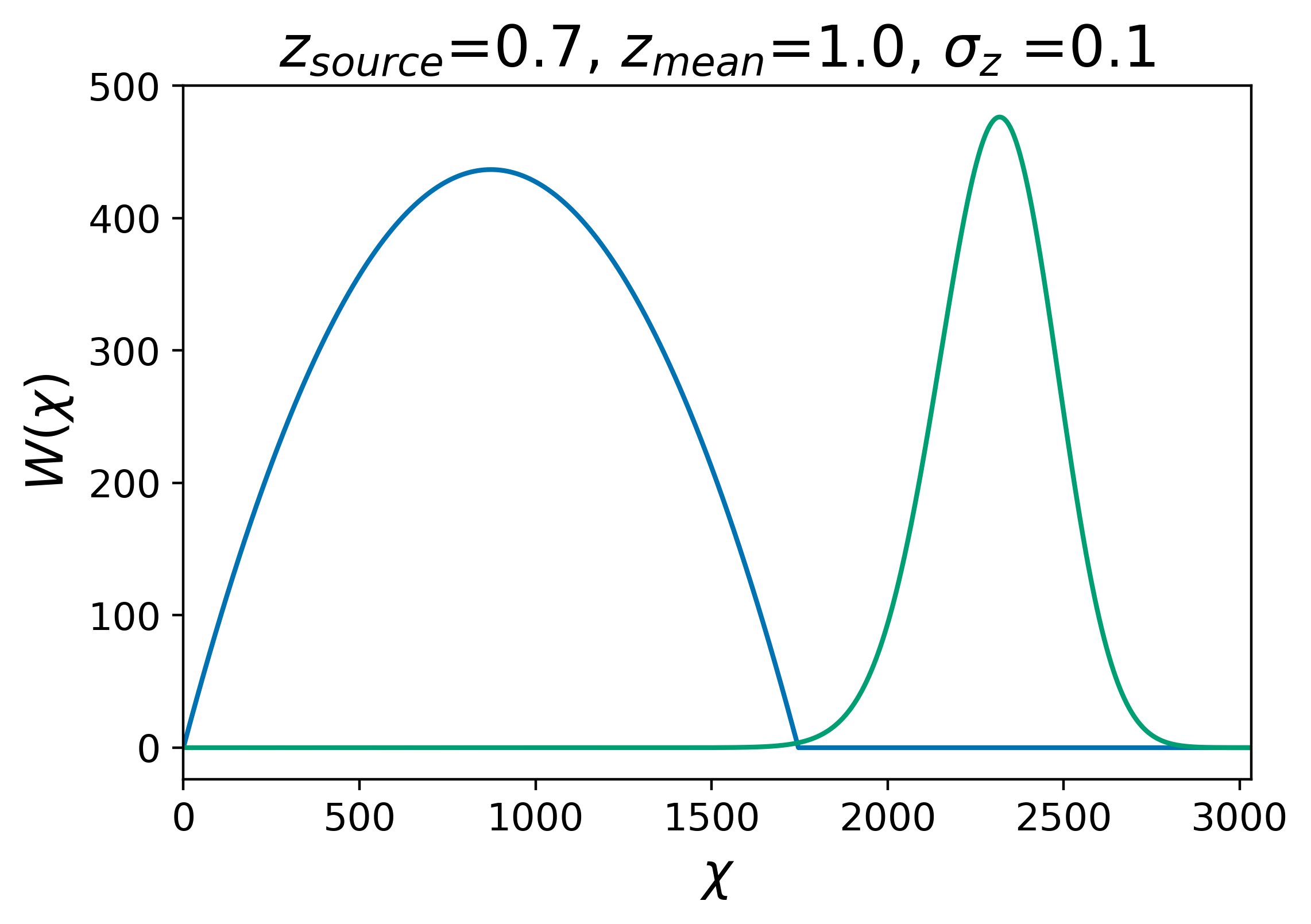}
\end{center}
\caption{\label{fig:RedshiftKernelsCross} Combinations of lensing kernels $W_\kappa(\chi,\chi_s)$ (blue) and galaxy distributions $W_g(\chi)$ (green) for which we evaluate cross correlations and their lensing corrections in this section. Note that we assume a single source redshift and no source redshift distribution. This choice should not affect the relative importance of lensing corrections significantly. The configuration in the first row was also used in Section~\ref{sec:limber} and we see that dropping the Limber approximation has not changed the correction significantly.}
\end{figure}
Results of the evaluations are shown in Figure~\ref{fig:CrossCorrectionnoLimber}. In the first column we show the cross-correlation signal along with the correction due to magnification bias for different values of the slope \say{$s$}. We point out that for all three examples, magnification bias plays an important role and we will get back to this in the next section. 
In the last column we plot the individual post Born correction terms (where we already sum the cancelling terms with the trick introduced in the previous section). The additional terms that vanish in Limber approximation are of comparable magnitude as the residuals of the Limber terms after cancellation. The relative importance of the terms varies with different combinations of source redshifts and galaxy kernels. We find a cancellation between another pair of terms (Eqs.~\ref{eq:22B} and~\ref{eq:31b}), which is very efficient for $z_s\gg z_\mean$, but disappears for $z_s < z_\mean$. In the last example, a term (Eq.~\ref{eq:13b}) that is zero under Limber and strongly suppressed for $z_s\gg z_\mean$ becomes similarly important as the Limber terms.
In the last column of Figure~\ref{fig:CrossCorrectionnoLimber}, we compare the sum of all of the above correction terms to the signal (for different amounts of magnification bias). The contribution is well below $0.01\%$ for all multipoles considered ($2<L<1000$). Extrapolating the curves seem to suggest that the relative correction decreases for multipoles $L>1000$, however, on these scales also non-linear corrections to the power spectrum become important and the employed line-of-sight integration decreases in precision, so a simple extrapolation might be misleading. 
\subsection{Additional Jacobian terms}
\label{sec:additional_terms}
So far we have ignored any higher order terms that involve the lensing Jacobian. These are, however, numerous and need to be taken into account for a full and consistent treatment. In this section we provide an overview of the full set of possible correction terms and make the case for their evaluation with simulations.

For simplicity, we first assume $s=0$, such that $J^{1-2.5s}=J$. With this simplification, we get the following Jacobian expansion terms up to third order
\begin{align}
J=& 1-2\kappa+\kappa^2-|\gamma|^2=1-2\[\kappa^{(1)}+\kappa^{(2)}
+\kappa^{(3)}\]\\
& +\[\kappa^{(1)}+\kappa^{(2)}\]^2-\[|\gamma|^{(1)}+|\gamma|^{(2)}\]^2 +\mathcal{O}(\kappa^{(4)}),
\end{align}
and identify
\begin{align}
J^{(0)}& = 1 \\
J^{(1)}& = -2 \kappa^{(1)}\\
J^{(2)}& = -2 \kappa^{(2)}+\[\kappa^{(1)}\]^2-\[|\gamma|^{(1)}\]^2\\
J^{(3)}& = -2 \kappa^{(3)}+2 \kappa^{(1)}\kappa^{(2)}-2 |\gamma|^{(1)}|\gamma|^{(2)}.
\end{align}
Using this notation we list all correction terms to the cross correlation that involve the lensing Jacobian (up to fourth order) in Table~\ref{tab:table2}.
\begin{table}[ht!]
  \begin{center}
    \begin{tabular}{l|l|l}
      order of Jacobian & possible combinations & evaluated in this work \\
      \hline
      $J^{(0)}$ & $\delta^{(1)}\kappa^{(3)},\,  \delta^{(3)}\kappa^{(1)},\, \delta^{(2)}\kappa^{(2)}$ & \checked, \checked, \checked\\
      \hline
      $J^{(1)}$ & $\kappa^{(1)},\, \kappa^{(3)},\, \delta^{(1)}{\kappa^{(2)}}^*,\, \delta^{(2)}{\kappa^{(1)}}^*$ & \checked, \texttimes,  \texttimes, \texttimes\\
      \hline
      $J^{(2)}$ & $\kappa^{(2)},\, \delta^{(1)}{\kappa^{(1)}}^*$ & \texttimes, \texttimes \\
      \hline
      $J^{(3)}$ & $\kappa^{(1)}$ & \texttimes \\
    \end{tabular}
    \caption{\label{tab:table2} Combination of higher order terms that enter the cross correlation up to fourth order in the density/deflection field (cp. Eq.~\ref{eq:gal_exp}). Note that $\kappa^{(3)}=\kappa^{(3a)}+\kappa^{(3b)}$ and similarly for $\delta^{(3)}$. We omit bispectrum terms. A star indicates that the expression for this term is given in Appendix~\ref{app:jacobian_terms}. }
  \end{center}
\end{table}
We provide expression for terms marked with a star in Appendix~\ref{app:jacobian_terms} and point out that we expect an efficient cancellation between the pure lensing terms $J^{(1)}\kappa^{(3)}+J^{(3)}\kappa^{(1)}+J^{(2)}\kappa^{(2)}\approx 0$.

Allowing an arbitrary amount of magnification bias ($s\neq 0 $) introduces another expansion on top of the terms listed listed above. Schematically,
\begin{equation}
    (1+x)^\beta = 1+ \beta x + \frac{1}{2} (\beta-1) \beta x^2 +\frac{1}{6} (\beta-2)(\beta-1)\beta x^3 +\mathcal{O}(x^4).
\end{equation}
\begin{figure}
\begin{center}
\includegraphics[width=0.45\textwidth]{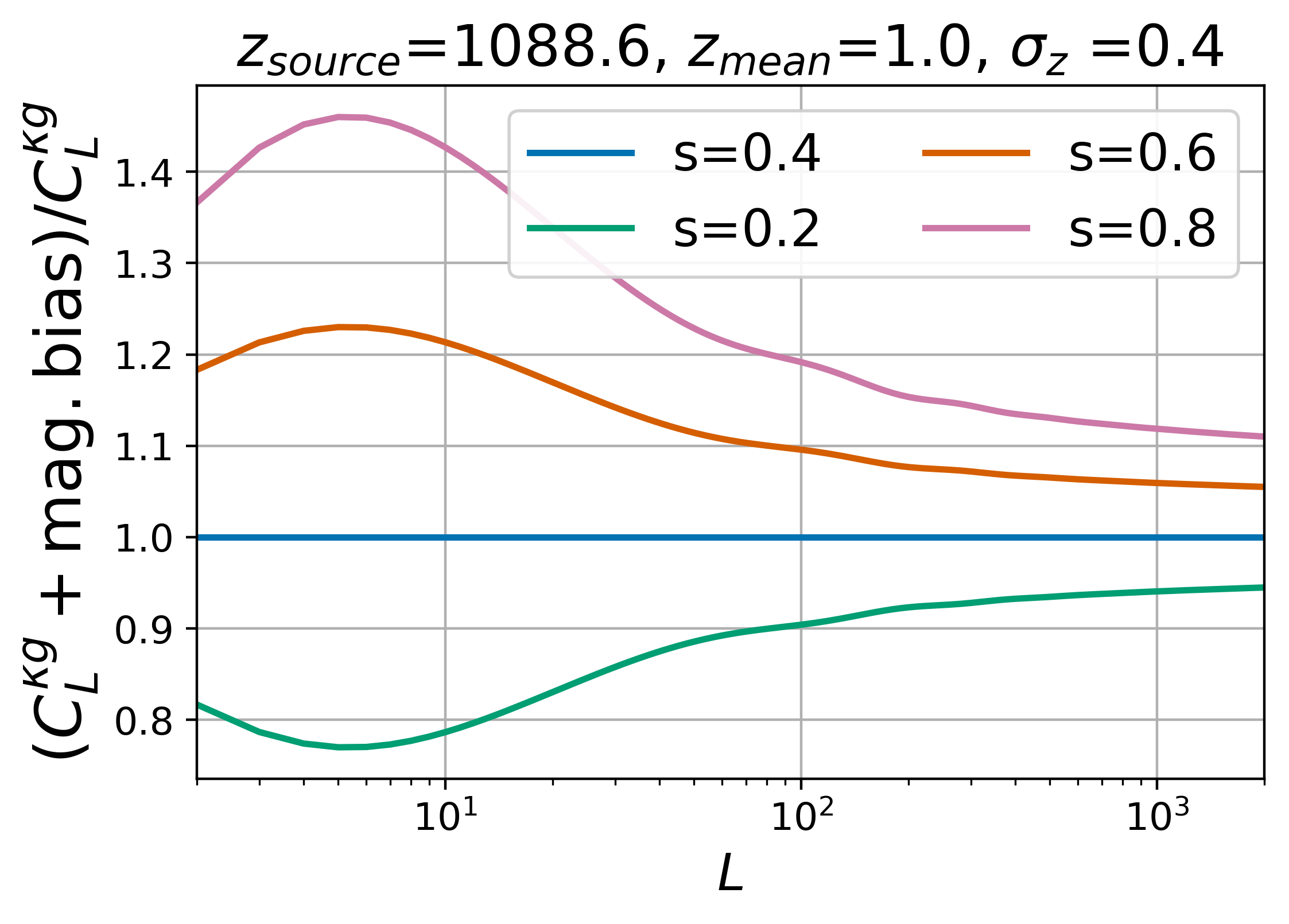}
\includegraphics[width=0.45\textwidth]{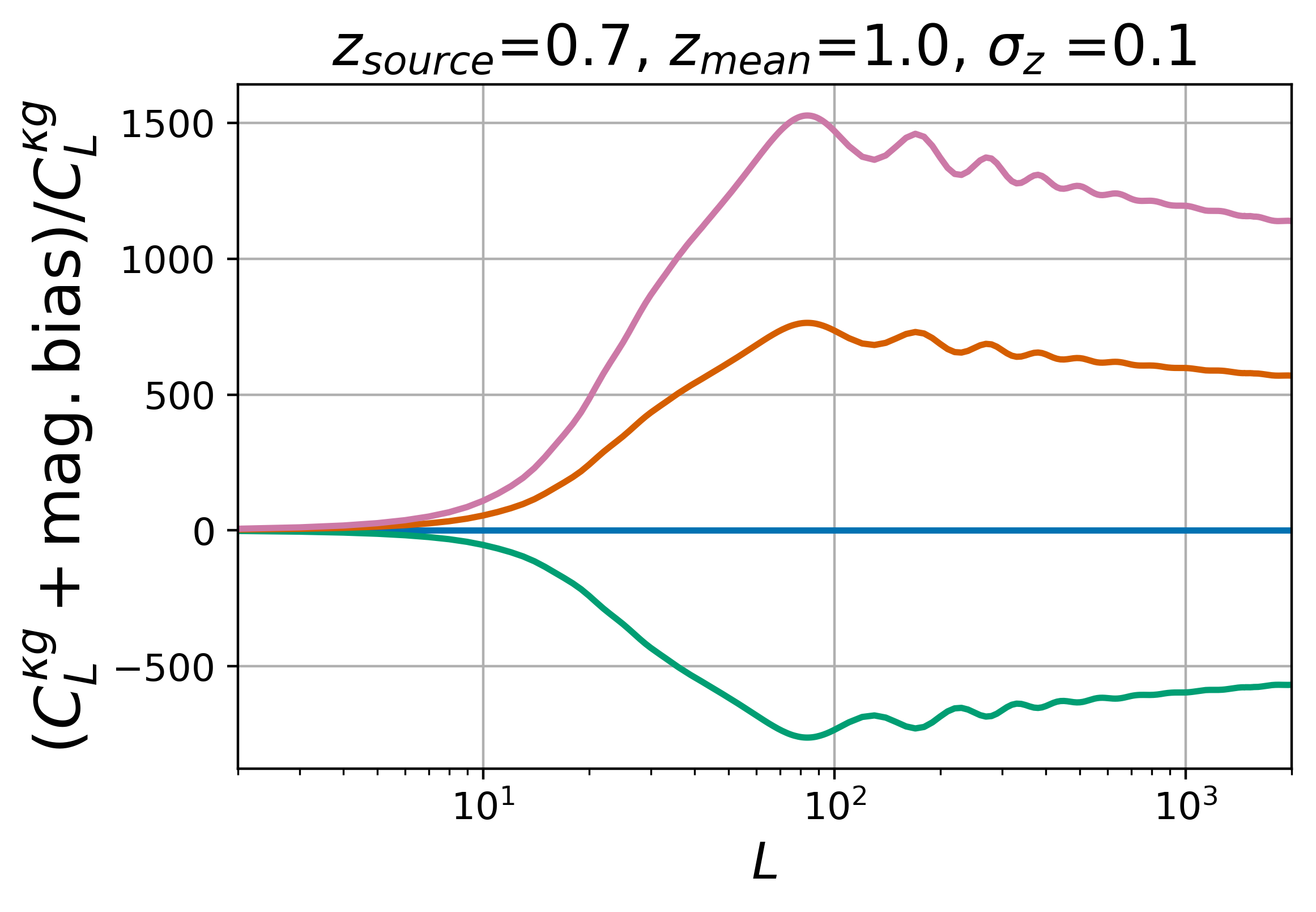}
\end{center}
\caption{\label{fig:magbiascross} Importance of magnification bias for different redshift configurations and number count slopes, $s$. Formally, magnification bias is of the same order as the signal itself (left plot). For unconventional configurations it can even dominate the signal (right plot).}
\end{figure}
In Fig.~\ref{fig:magbiascross} we show the relative change of the signal after including the leading order magnification bias correction to illustrate that it can easily be of the same order of magnitude as the signal itself (and should really be considered as part of the signal). This in turn suggests that correction terms involving the lensing Jacobian should be similarly important as the terms evaluated in the previous section. 
Given the high number of contributing correction terms after allowing for arbitrary magnifications, these terms could add up to a significant correction. Due to their sheer number and involved evaluation, we suggest the use of ray-traced lensing simulation to estimate their total effect. 
\section{Lensing corrections to the galaxy auto power spectrum}
\label{sec:gal_gal}
We now turn to evaluating the same set of terms as in Section~\ref{sec:no_limber}, but for the correlation of galaxy number counts in different redshift windows.
The corresponding expressions are
\begin{align}
\nonumber
C_{22}^{(g g)}(L)  = &  \int \frac{\d^2 \vl}{(2 \pi)^2} \[\vl \cdot \(\VL-\vl\)\]^2 \int_0^{\chi_\max} \d \chi_1 \, W^{(1)}_g(\chi_1)  \int_0^{\chi_\max} \d \chi_2 \, W^{(2)}_g(\chi_2) \\
\label{22auto}
& \[ C_l^{\delta \delta} (\chi_1, \chi_2) C_{|\VL-\vl|}^{\phi\phi}(\chi_1,\chi_2)+ C_l^{\delta \phi}(\chi_1, \chi_2) C_{|\VL-\vl|}^{\delta\phi}(\chi_2,\chi_1)\]
\end{align}
The above term is symmetric in the redshift kernel and equivalent to Eqs.~\ref{eq:22A}-\ref{eq:22B} in the cross correlation. Similarly,
\begin{align}
\label{13aauto}
C_{13/31a}^{(g g)}(L)  = & - \frac{1}{2}\int \frac{\d^2 \vl}{(2 \pi)^2} \[\vl \cdot \VL\]^2  \\
\nonumber
& \int_0^{\chi_\max} \d \chi_1 \, W^{(1)}_g(\chi_1) \int_0^{\chi_\max} \d \chi_2 \, W^{(2)}_g(\chi_2) C_L^{\delta \delta} (\chi_1, \chi_2) C_{l}^{\phi\phi}(\chi_1)\\
\nonumber
&+[(W^{(1)},\chi_1)\leftrightarrow (W^{(2)},\chi_2)]
\end{align}
is the equivalent of Eqs.~\ref{eq:31aA} and~\ref{eq:13a}. An equivalent to Eq.~\ref{eq:31aB} seems to be absent (parity) due to the different derivative structure.

The remaining term vanishes in the Limber approximation,
\begin{align}
\nonumber
C_{13/31b}^{(g g)}(L)  = & 2 \int \frac{\d^2 \vl}{(2 \pi)^2} \[\vl \cdot \VL\]^2 \int_0^{\chi_\max} \d \chi_1 \, W^{(1)}_g(\chi_1) \int_0^{\chi_\max} \d \chi_2 \, W^{(2)}_g(\chi_2) \\
\nonumber
& \int_0^{\chi_1} \d \chi_1' \, W(\chi_1',\chi_1) C_l^{\delta \phi} (\chi_1, \chi_1') C_{L}^{\Psi\delta}(\chi_1',\chi_2)\\
\label{31bauto}
& +[(W^{(1)},\chi_1)\leftrightarrow (W^{(2)},\chi_2)].
\end{align}
The correction terms to the galaxy-galaxy correlation function are similar to those to the cross correlation. However, there is no direct correspondence (i.e. we cannot simply replace the lensing kernel by a galaxy kernel and $\psi$ by $\delta$ to go from the expressions for the cross correlation to the auto correlation). The reason for this is the different derivative structure, which results in different terms vanishing due to parity. We evaluate the corrections on the galaxy correlation in the same fashion as the corrections on the cross correlations in the previous section.

\begin{figure}
\begin{center}
\includegraphics[width=0.32\textwidth]{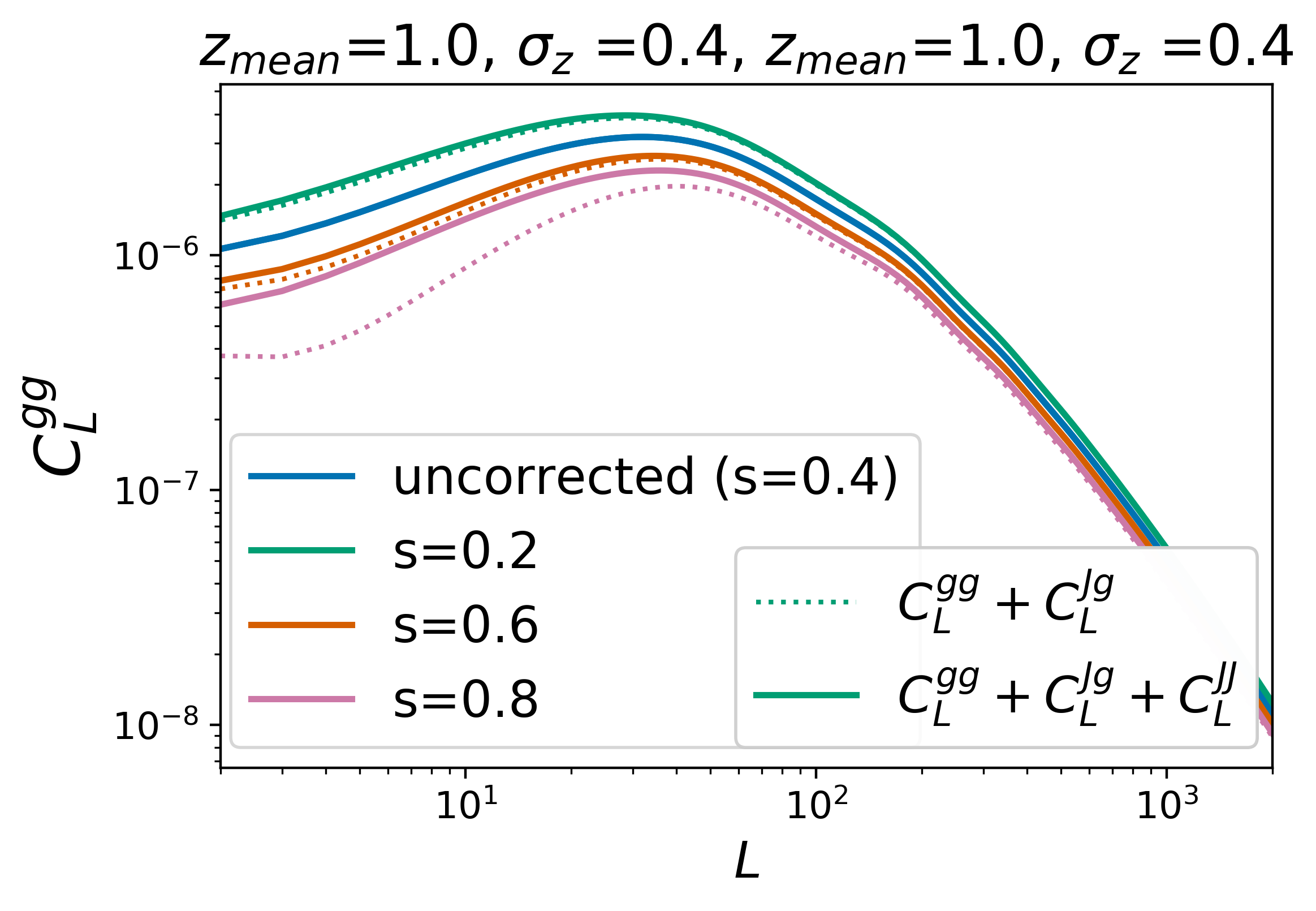}
\includegraphics[width=0.32\textwidth]{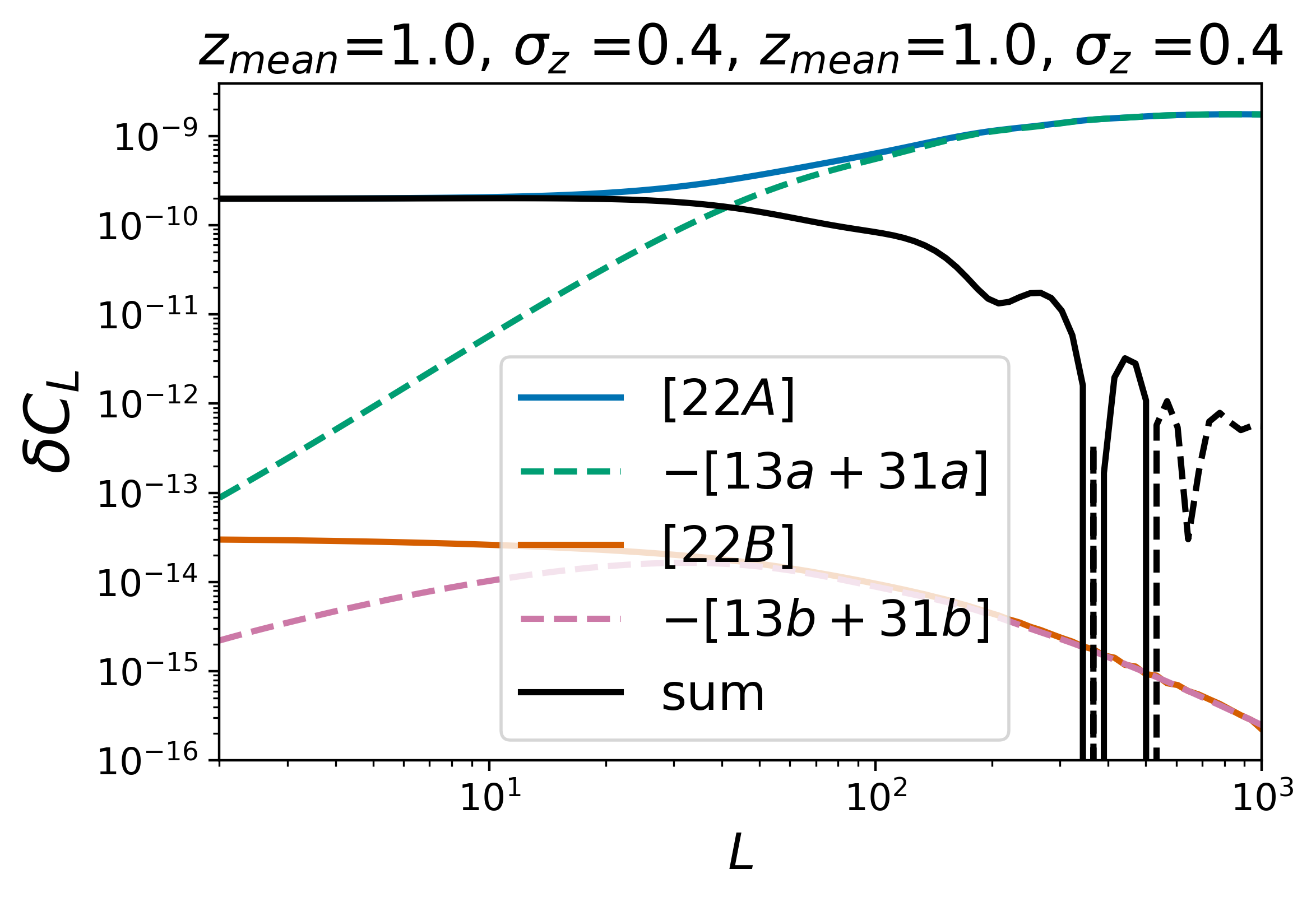}
\includegraphics[width=0.32\textwidth]{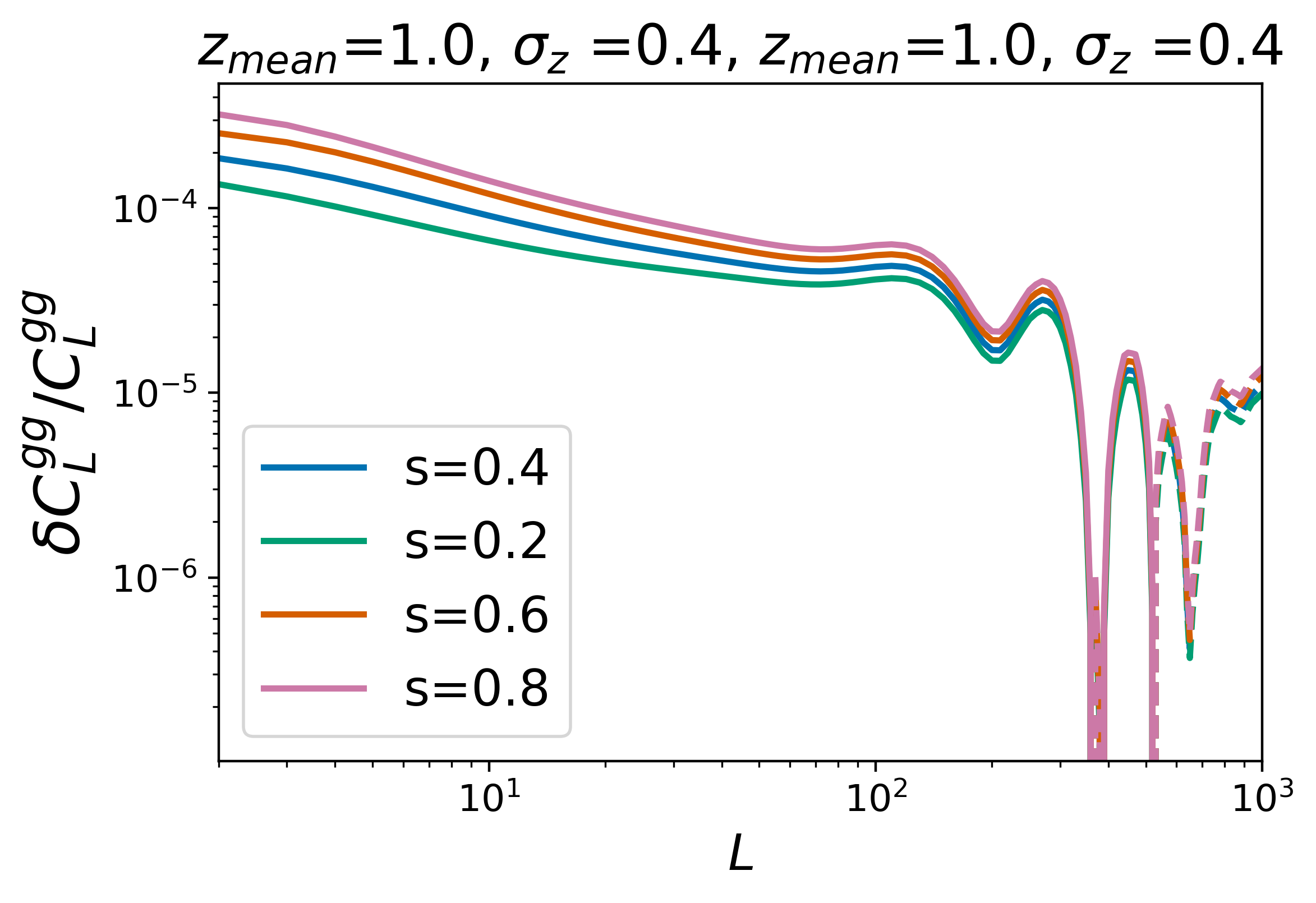}
\includegraphics[width=0.32\textwidth]{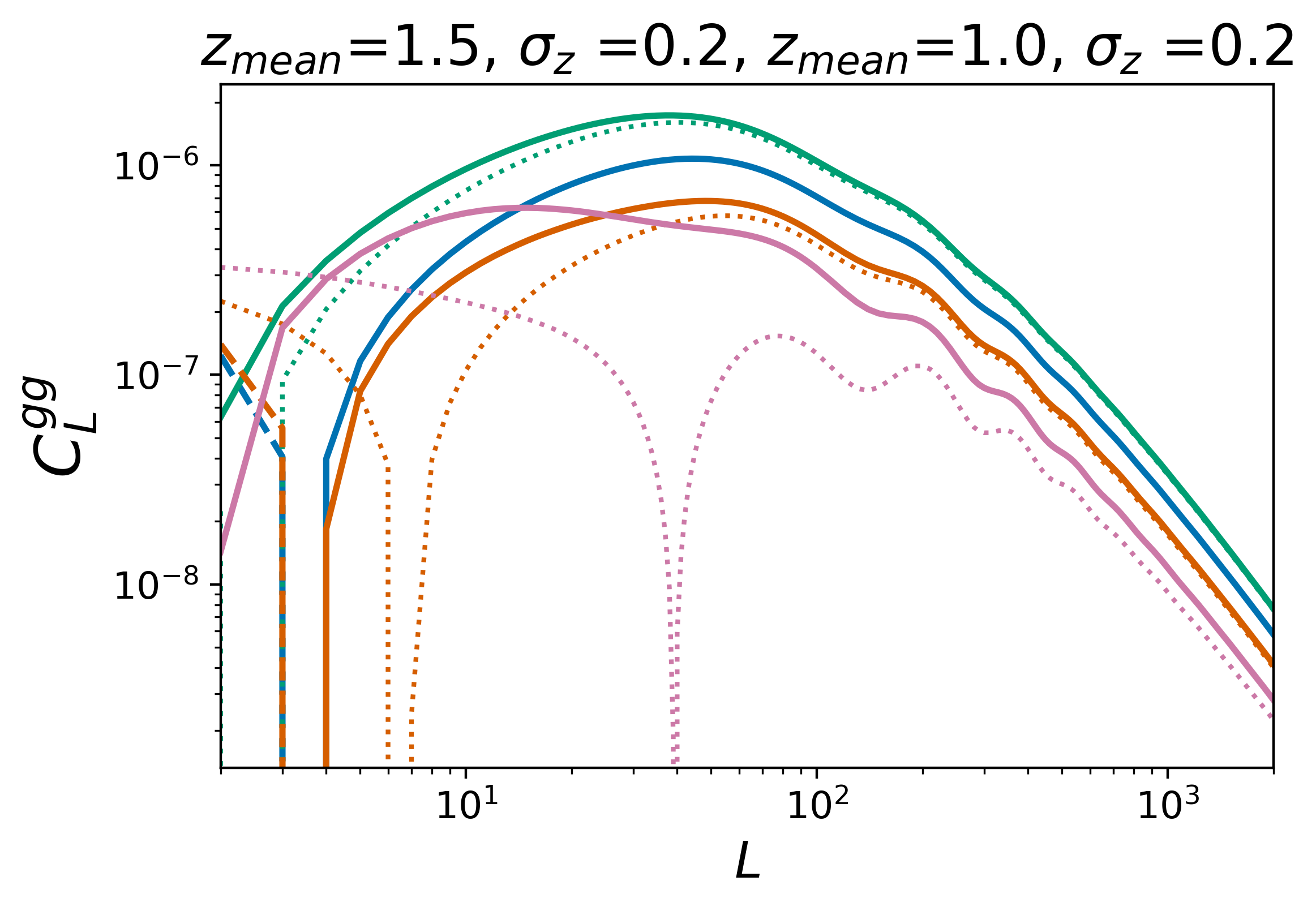}
\includegraphics[width=0.32\textwidth]{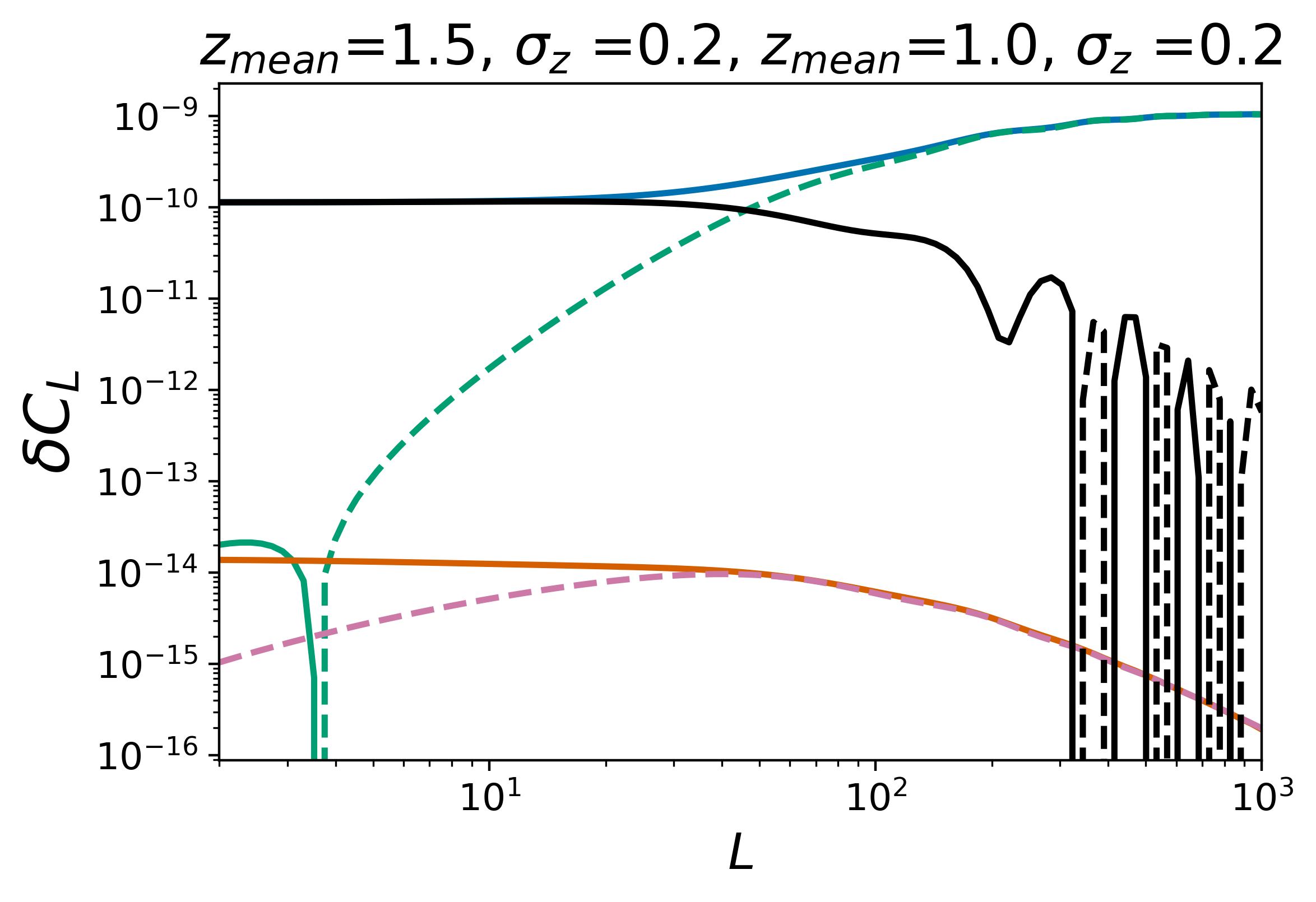}
\includegraphics[width=0.32\textwidth]{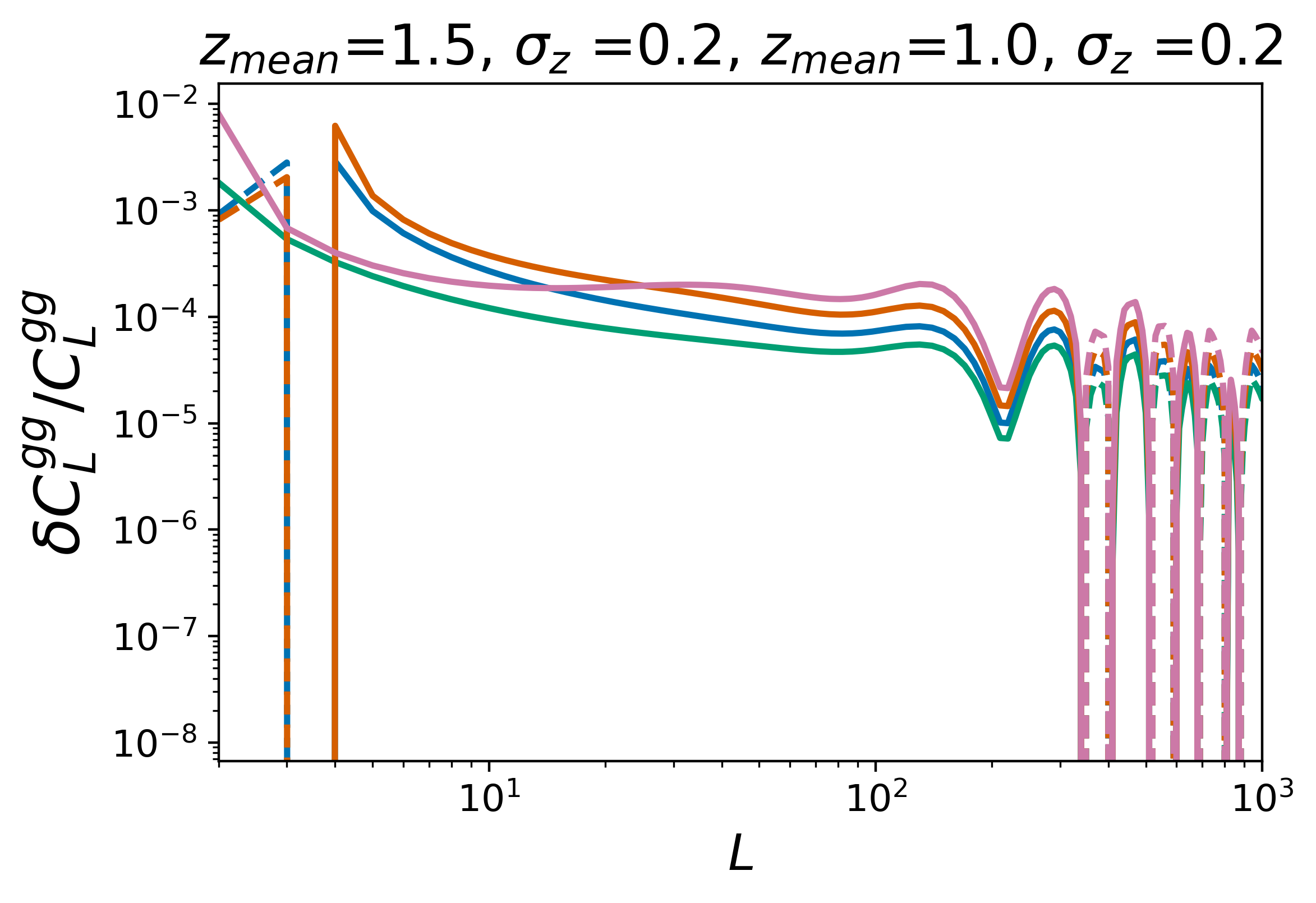} 
\includegraphics[width=0.32\textwidth]{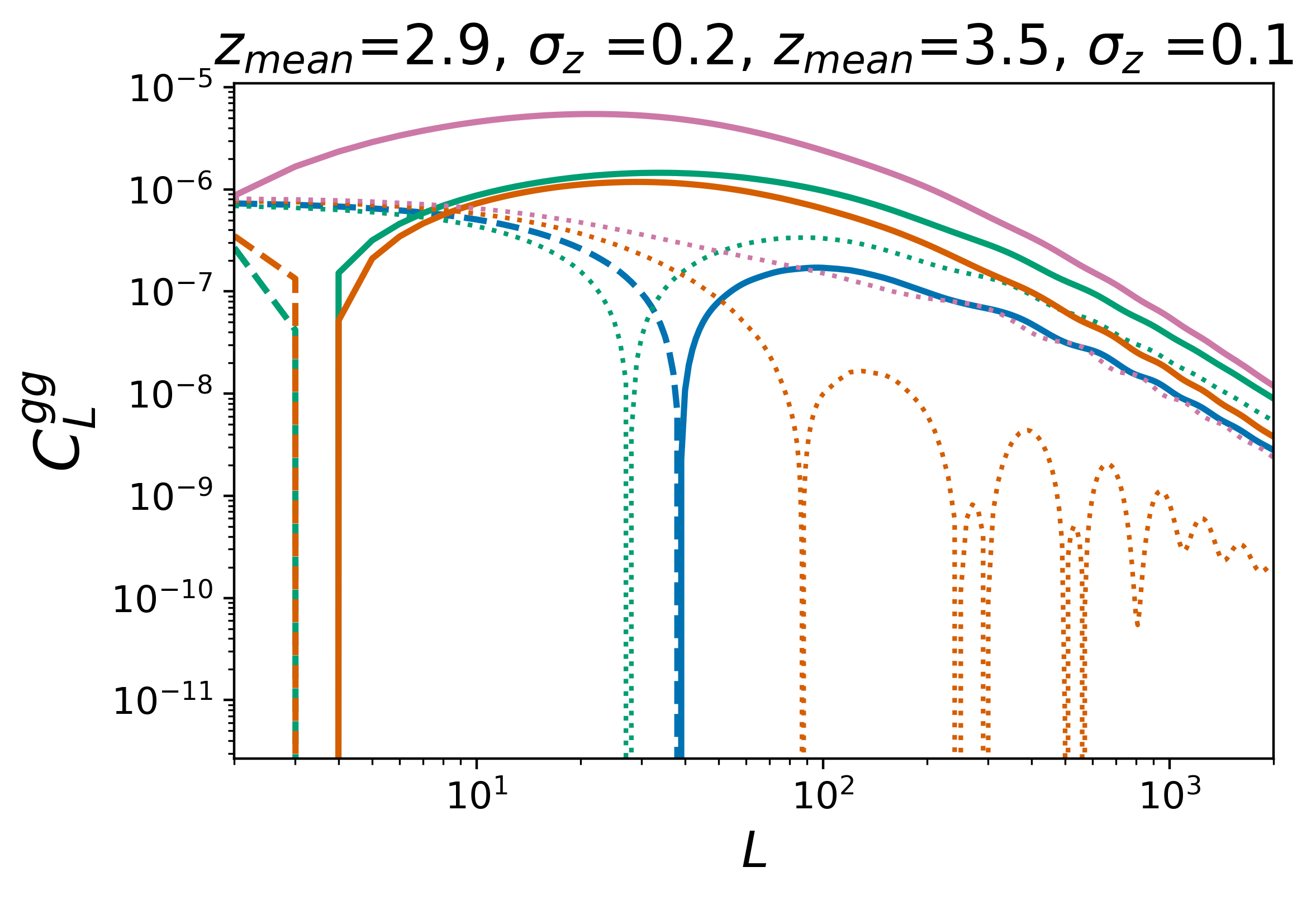}
\includegraphics[width=0.32\textwidth]{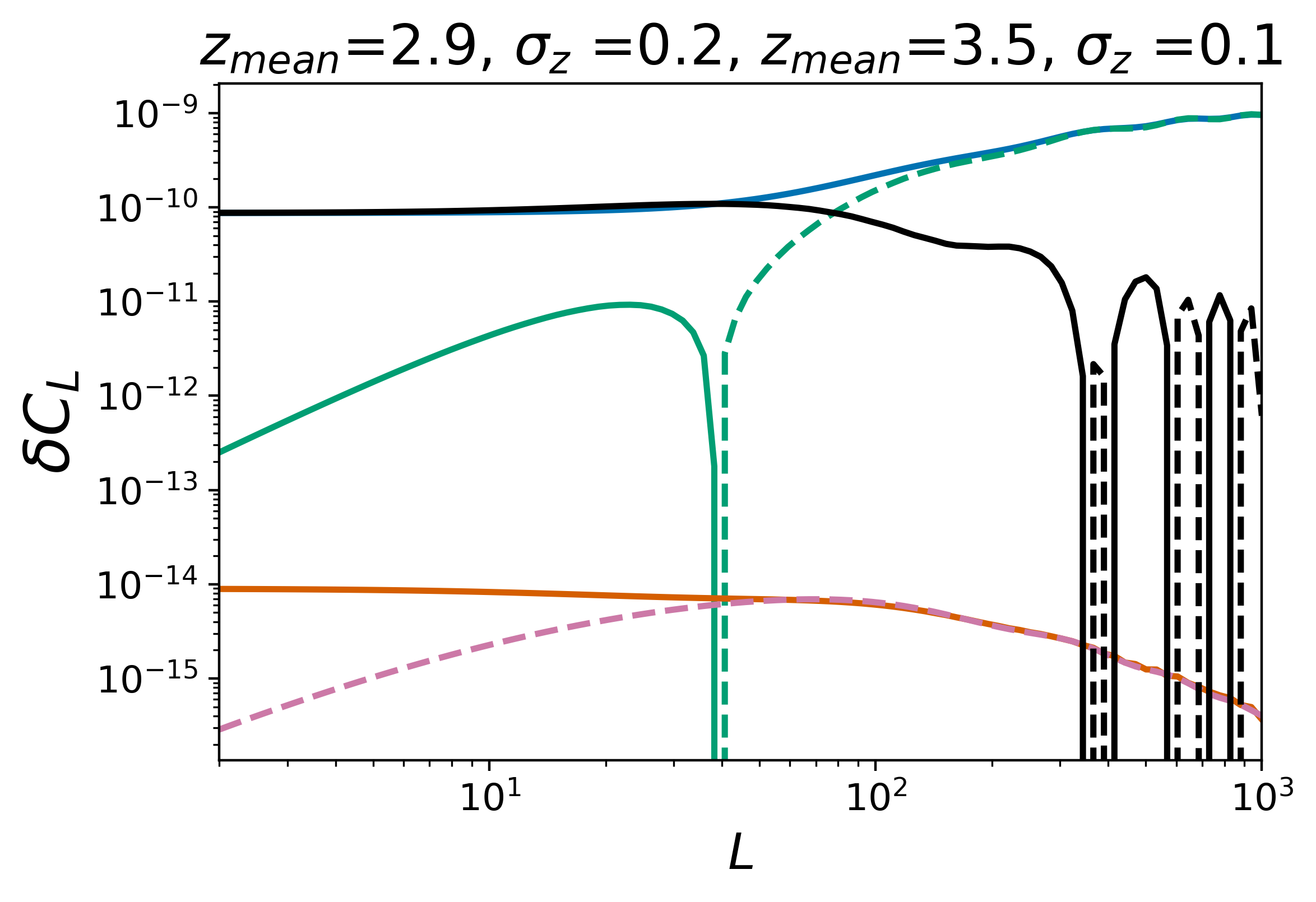}
\includegraphics[width=0.32\textwidth]{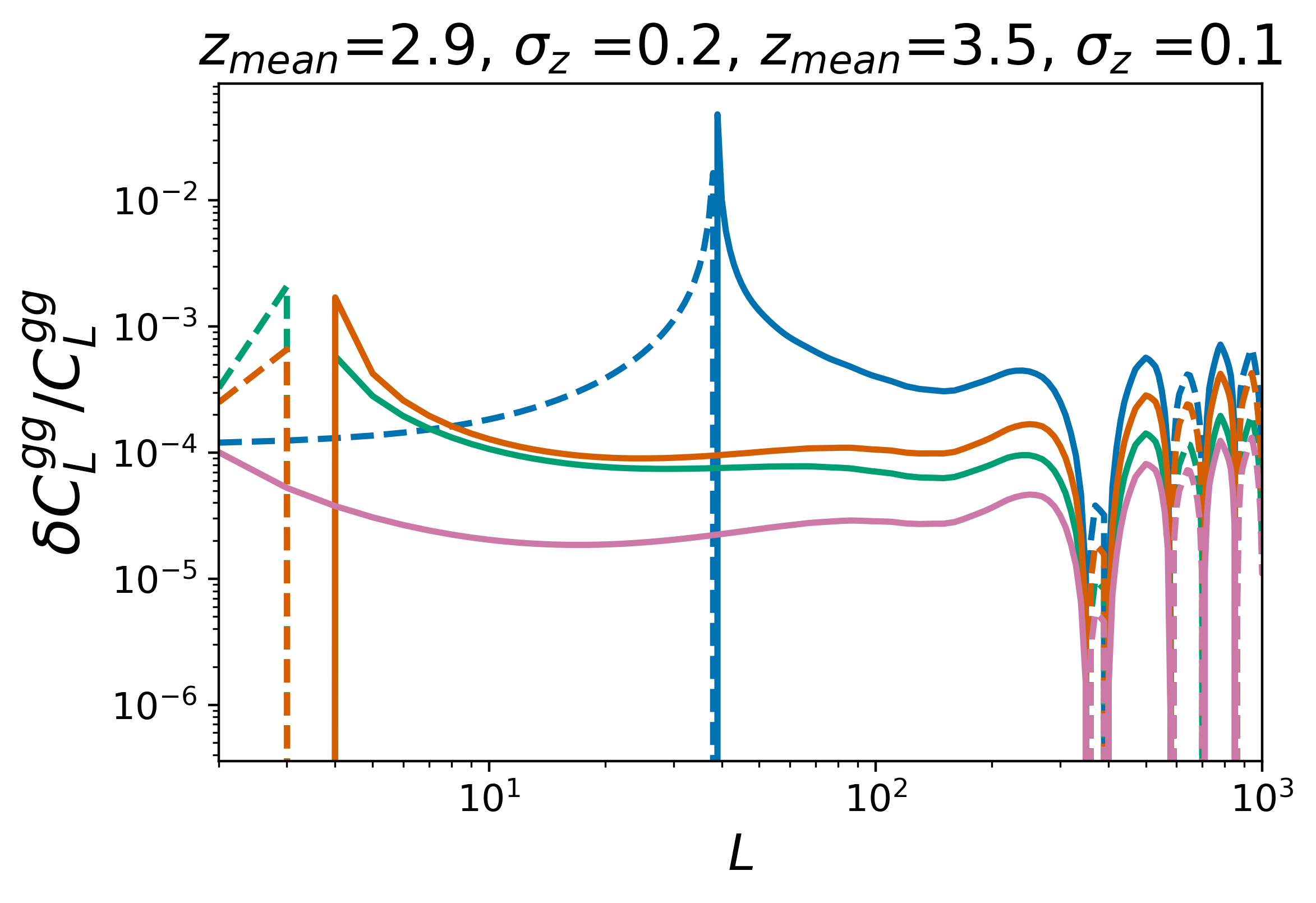}
\includegraphics[width=0.32\textwidth]{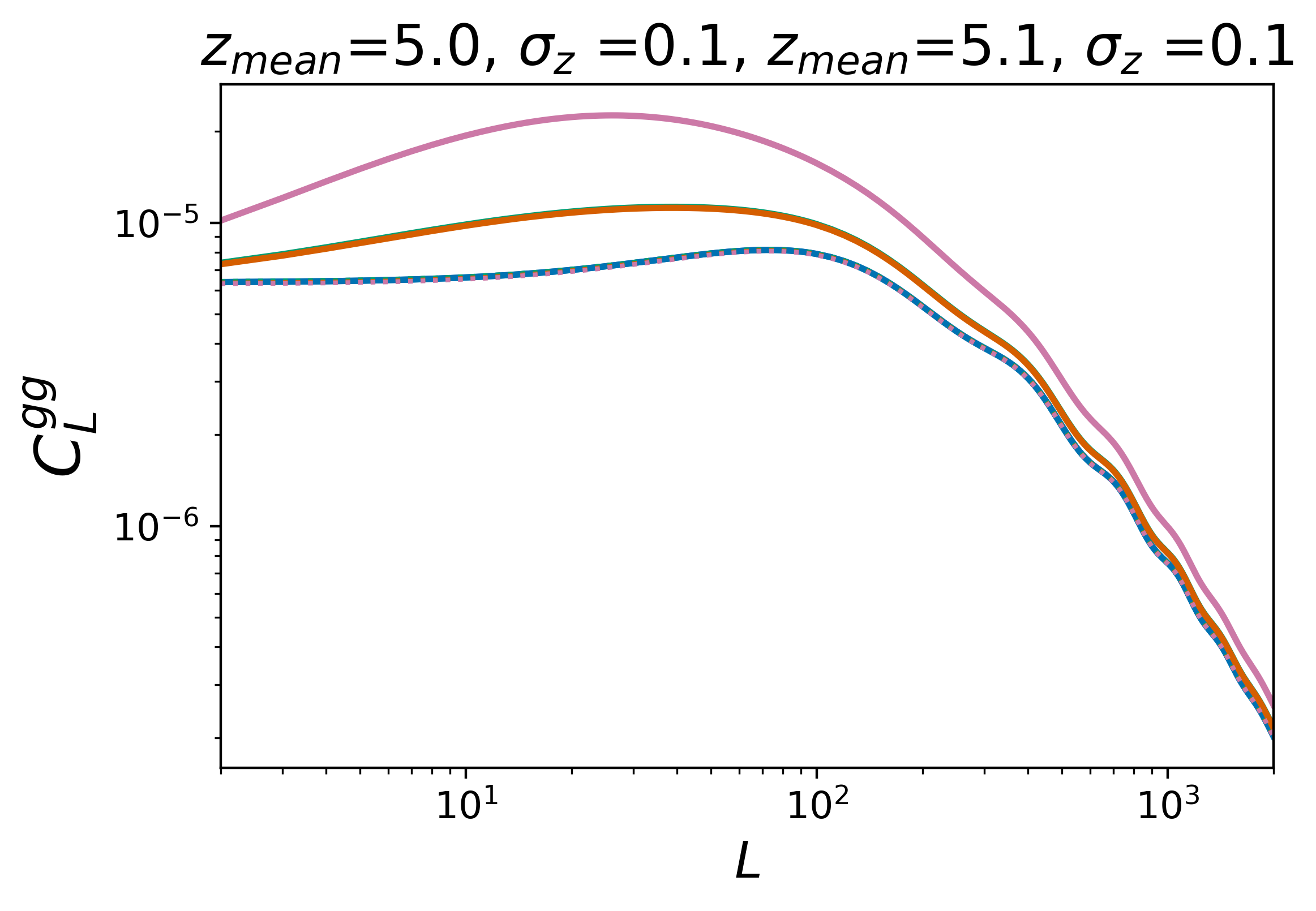}
\includegraphics[width=0.32\textwidth]{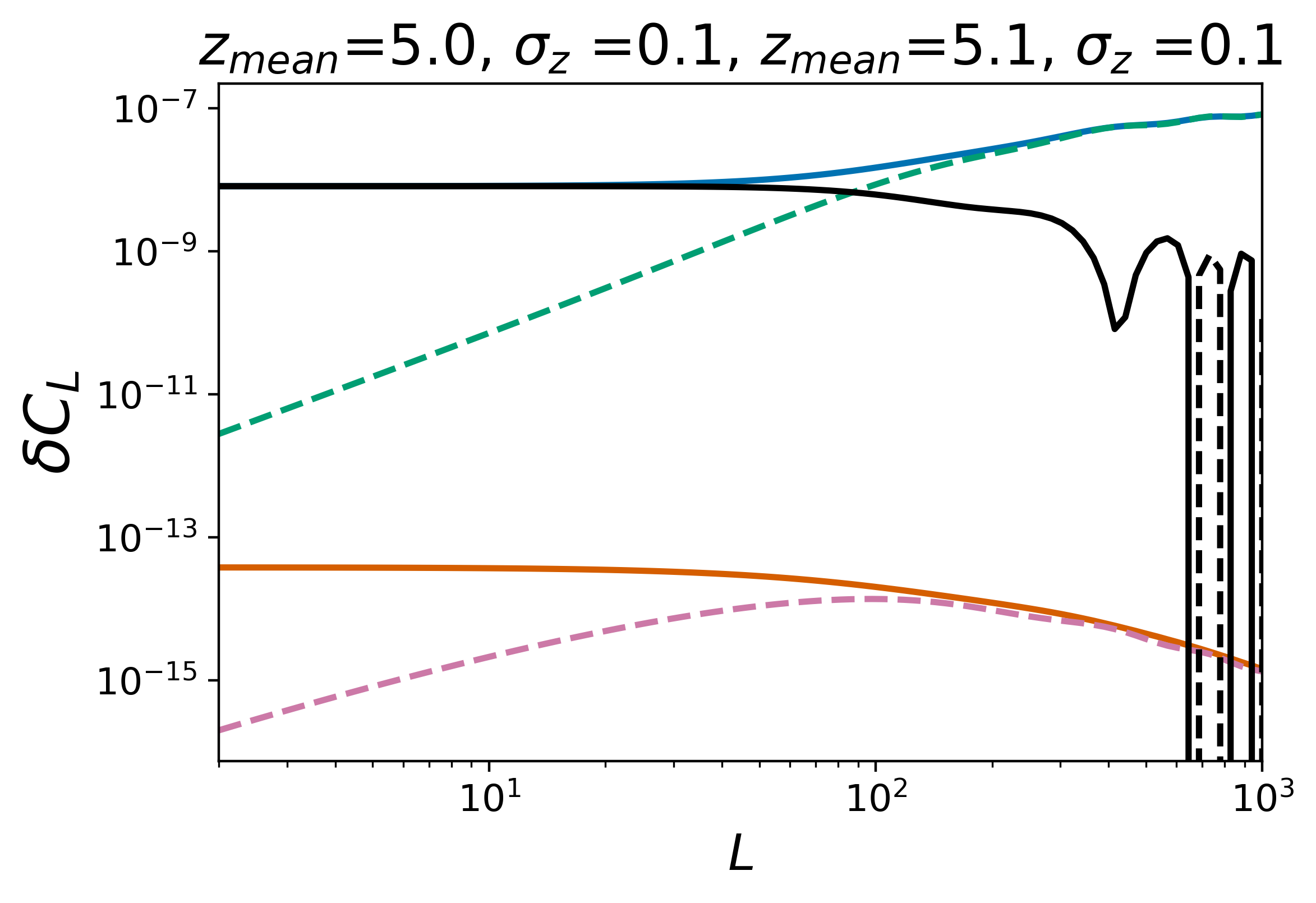}
\includegraphics[width=0.32\textwidth]{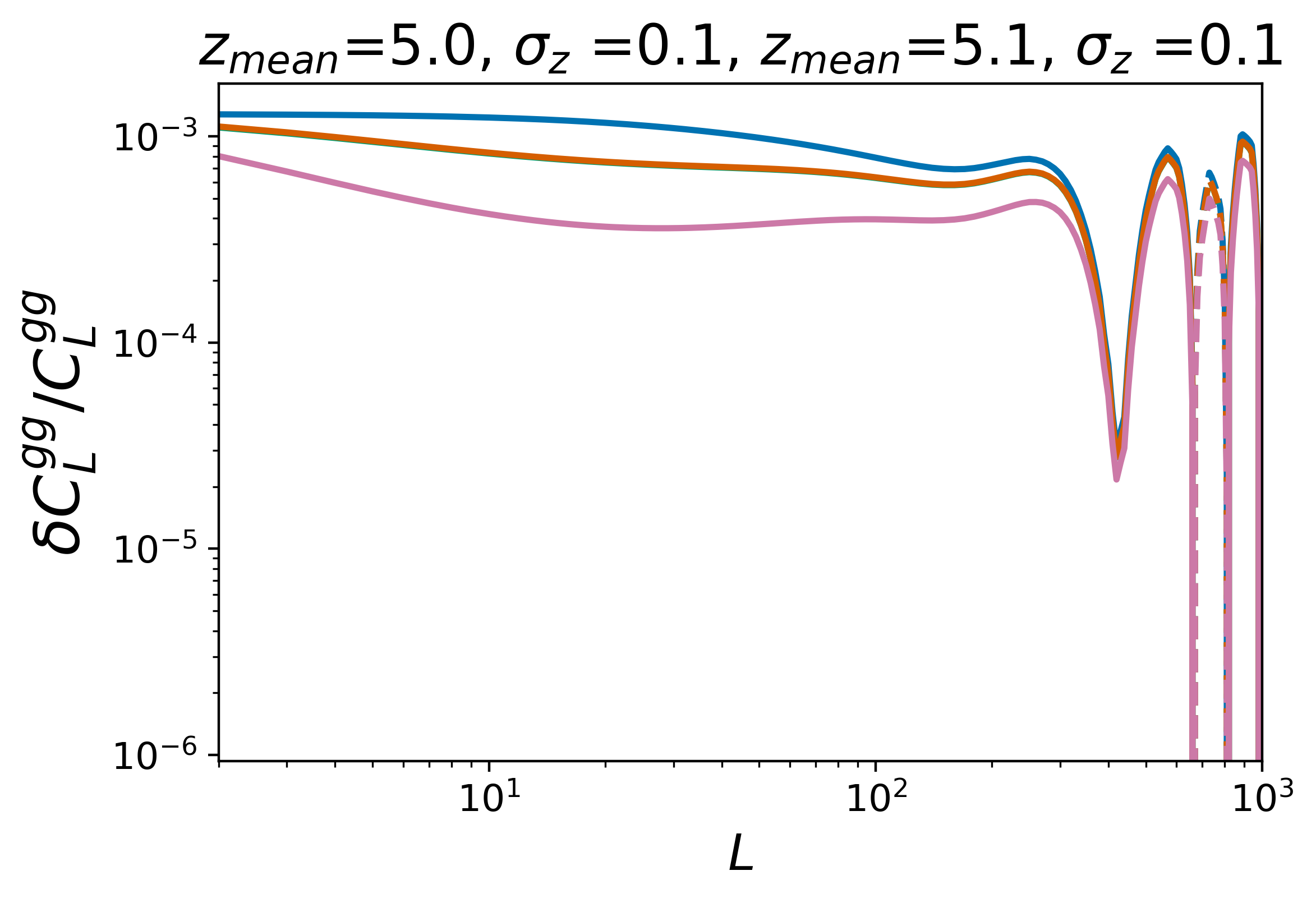}
\end{center}
\caption{\label{fig:AutoCorrectionnoLimber} Lensing corrections (without Jacobian terms) on $C_L^{gg}$ for different combinations of Gaussian redshift distributions: In the first column we plot the signal without magnification bias (blue) and corrected by the two arising leading order magnification bias terms (dotted lines show the effect of correcting for one of the terms only). In the middle column we plot the lensing correction terms (Eqs.~\ref{22auto}-\ref{13aauto}, their sum is indicated in black. On small scales, we find the same cancellation as for $C_L^{\kappa g}$ and $C_L^{\kappa \kappa}$. In the last column, we plot the ratio of these corrections to the signal. The correction is biggest on large scales (where also the signal is smaller), and below $<1\%$. The gap in some of the plots is due to our L-sampling; we sample only at integer numbers. }
\end{figure}

As in the previous section, in addition to lensing corrections, 
we also have to include magnification bias~\citep{magbias3, magbias4}, which has two contributions, the galaxy-convergence correlation
\begin{align}
C_{1J1}^{(g g)}(L)  &= \frac{1}{2} 5(s-0.4) L^2 \int_0^{\chi_\max} \d \chi_1 \, W^{(1)}_g(\chi_1) \int_0^{\chi_\max} \d \chi_2 \, W^{(2)}_g(\chi_2) C_l^{\delta \phi} (\chi_1, \chi_2)
\label{magbias_gg1}
\end{align}
and the convergence-convergence correlation
\begin{align}
C_{J1J1}^{(g g)}(L)  &= \[\frac{1}{2} 5(s-0.4) L^2\]^2 \int_0^{\chi_\max} \d \chi_1 \, W^{(1)}_g(\chi_1) \int_0^{\chi_\max} \d \chi_2 \, W^{(2)}_g(\chi_2) C_l^{\phi \phi} (\chi_1, \chi_2).
\label{magbias_gg2}
\end{align}
Results for different combination of redshift distributions are shown in Fig~\ref{fig:AutoCorrectionnoLimber}. To illustrate the effect of magnification bias, we plot in the first column the correlation function without correction (in blue), the correlation function after correcting for the first term (dotted lines) and after correcting for both terms (other colored, solid lines). For the lensing corrections we find two very efficient cancellations on small scales, independent of the respective redshift distributions. One amongst the terms that are non-zero in Limber
\begin{equation}
[22A]+[31/13\mathrm{a}] \approx 0
\end{equation}
and the other amongst the terms that are zero in Limber
\begin{equation}
[22B]+[31/13\mathrm{b}] \approx 0.
\end{equation}
The first cancellation can be directly seen by comparing the first term in Eq.~\ref{22auto} with Eq.~\ref{13aauto} and taking the limit $L\gg l$. The second cancellation can be seen by noticing that $C_{L}^{\Psi\delta}(\chi_1',\chi_2)$ in Eq.~\ref{31bauto} peaks sharply at $\chi_1'=\chi_2$. Using this as a constraint (in the second term of eq.~\ref{31bauto}, correspondingly the constraint is $\chi_1'=\chi_1$), we see that we approximately recover the second term of Eq.~\ref{22auto}. 

The cancellations are depicted in the second column of Fig~\ref{fig:AutoCorrectionnoLimber}. In the last column we plot the relative size of the corrections to the signal. While being bigger than in the auto-correlation, the corrections still remain at the sub-percent level a result which is in agreement with estimates from ray-traced simulations~\citep{Beck2019}.

\begin{figure}
\begin{center}
\includegraphics[width=0.32\textwidth]{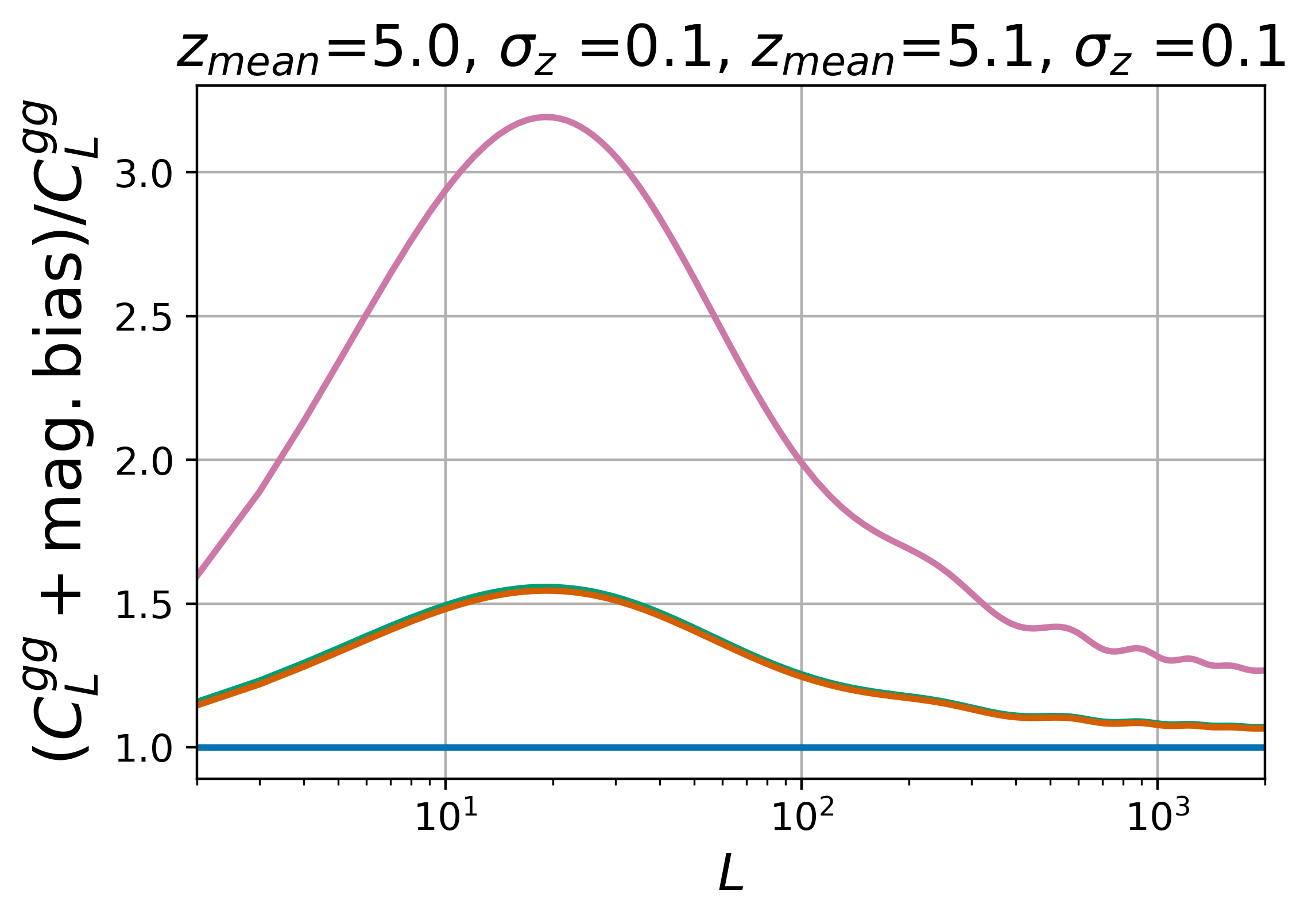}
\includegraphics[width=0.32\textwidth]{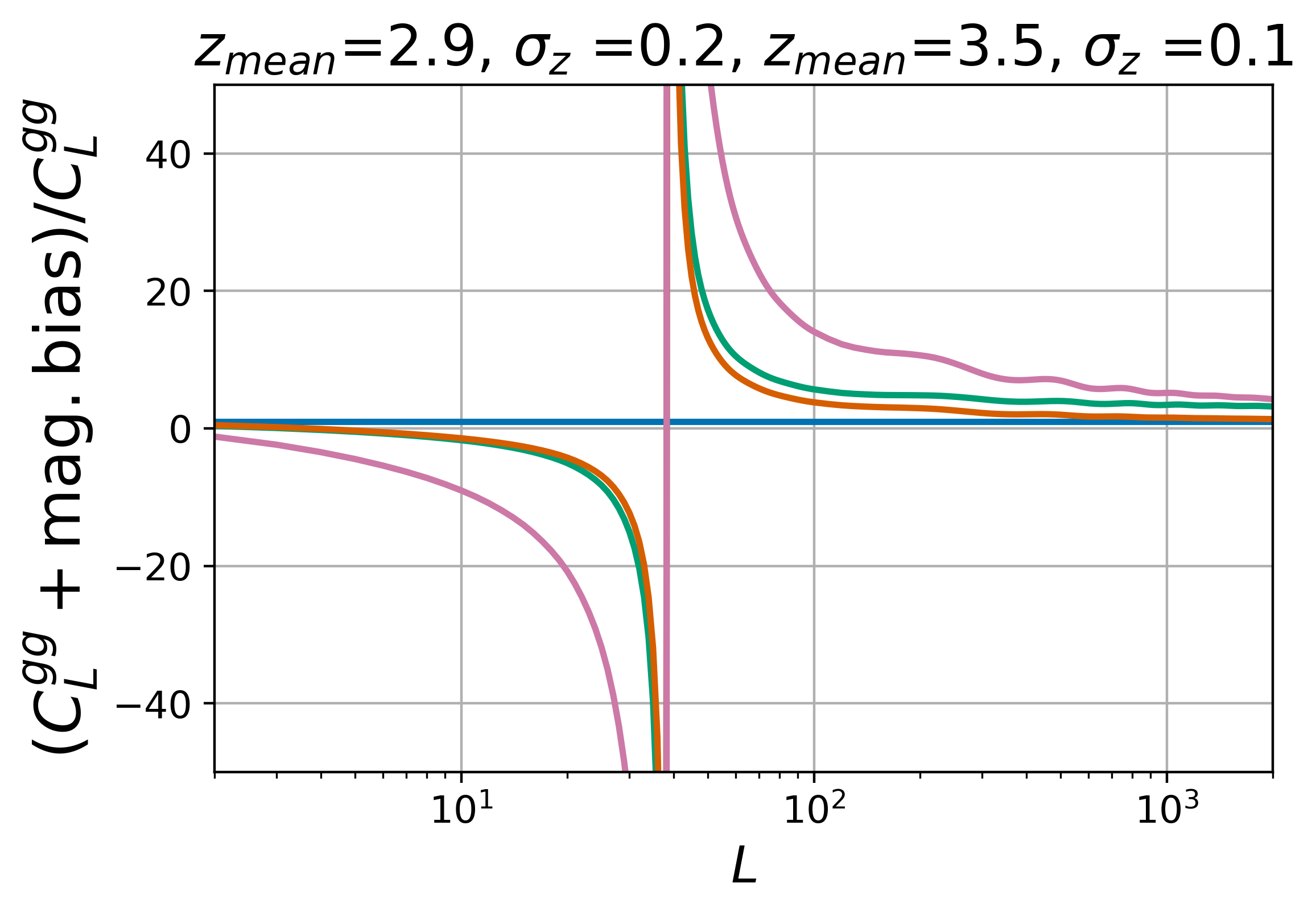}
\includegraphics[width=0.32\textwidth]{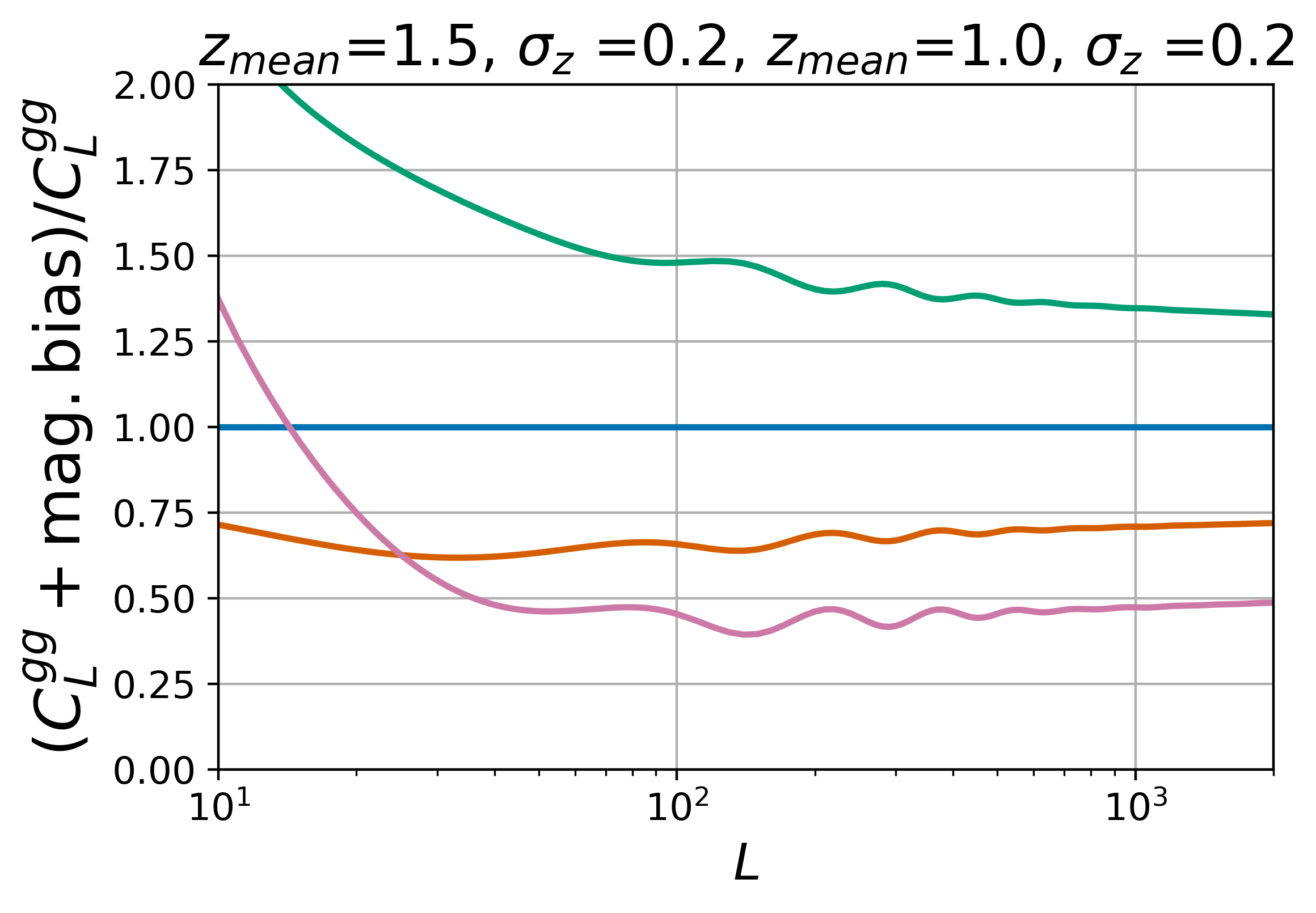}
\end{center}
\caption{\label{fig:magbiasauto} Illustrating the importance of magnification bias for galaxy number counts. We show here examples where it can change the signal by a factor of 2 or more. The importance of a correct modeling of the magnification bias for this observable was also recently pointed out by Ref.~\citep{2019arXiv190910539T}.}
\end{figure}

As in the previous section, we have again neglected the numerous terms that arise when allowing for higher order Jacobian terms. In Fig.~\ref{fig:magbiasauto} we plot the relative change of the signal when including magnification bias terms. The size of this change (in the examples up to a factor 2 and higher) suggests that higher order Jacobian terms should be taken into account in a full analysis.
\section{Conclusions}
\label{sec:conclusions}
In this work we studied the importance of lensing corrections for modelling several types of (cross-) correlations: correlating CMB lensing with galaxies , weak galaxy lensing with galaxies and galaxies with galaxies (possibly at different redshifts). Studies of this type are important given the required accuracy for modeling future measurements of cosmological parameters and the mass of neutrinos from these observables. 

Our approach does not make use of the Limber approximation and we show that terms that vanish in the Limber approximations can be of similar importance as terms that are non-zero in the Limber approximation.
We find efficient cancellations between higher order lensing correction terms, also when cross correlating fields at different redshifts and discuss the physical origin of these cancellations. 
We provide fast and accurate code (\href{https://github.com/VMBoehm/lensing-corrections}{\faGithub}~\citep{code}, which has undergone double checks at every stage) that can be used to reproduce our results and to estimate the size of lensing corrections for all aforementioned observables and their arbitrary redshift combinations.

Beyond the type of correction terms that also arise for lensing auto correlations we identify additional terms, caused by the lensing jacobian correction to galaxy number counts. We do not evaluate these terms but point out that their sheer number could add up to some significance. Future work should estimate the size of all these corrections with ray-traced simulations, as well as their impact on parameter estimates from these measurements.
\acknowledgments{We thank Simon Foreman, Emmanuel Schaan, Enea di Dio, Giulio Fabbian and Antony Lewis for useful comments on the draft. This research used resources of the National Energy Research Scientific Computing Center (NERSC), a U.S. Department of Energy Office of Science User Facility operated under Contract No. DE-AC02-05CH11231.}
\appendix
\section{Appendix: Derivation of higher order terms in weak lensing}
\label{app:lens_theory}
Weak lensing theory aims at relating an observed angular extent, $\vtheta$, of a an object (source) in the sky to the angular extent, $\vbeta$, it would have if there was no gravitational lensing (see Ref.~\citep{2001Bartelmann} for a review). By writing 
\beq
\label{simple_lensing}
\vbeta = \vtheta + \Delta \vtheta, 
\eeq
we define the (scaled) lensing deflection angle, $\Delta \vtheta$, which is the total deflection of a light ray generated by all lenses encountered along its geodesic. In weak lensing and in the Newtonian gauge, an expression for the deflection can be derived from the the relation between the transverse, comoving distance $\vx$ that separates two light rays at distance $\chi$ and the encountered metric perturbations, $\Psi$, that act as lenses
\beq
\label{eq:vx}
\vx(\vtheta, \chi) = \chi \vtheta - 2\int_0^{\chi}\d \chi' \(\chi-\chi'\)  \nabla_{\vx} \Psi\[\vx(\vtheta,\chi'),\chi' \]. 
\eeq
This expression is not closed, since every lens changes the photon's path and thus the metric perturbations that the photon will encounter next.  
If the lensing deflections are small, the separation, $\vx$, can be expanded into a leading order (no-lensing) contribution and higher-order lensing corrections,
\beq
\vx(\vtheta, \chi) = \vx^{(0)} +\Delta \vx = \vx^{(0)} + \Delta \vx^{(1)}+\Delta \vx^{(2)} + \mathcal{O}\[(\nabla \Psi)^3\].
\eeq
At lowest order there is no lensing ($\Delta \vtheta=0$) and for a source at distance $\chi$, we have $\vbeta = \vx / \chi$. The leading order contribution to $\vx$ is therefore 
\beq
\vx^{(0)}(\vtheta, \chi)= \vtheta \chi.
\eeq
To obtain the lensing terms, we expand the local deflection $\nabla \Psi$ in $\Delta \vx$,
\begin{equation}
    \nabla_{\vx} \Psi(\vx,\chi) =  \nabla_{\vx}\Psi(\vx,\chi)\rvert_{\vx =\vx^{(0)}}+ \nabla_{\vx} \nabla_{\vx_i} \Psi(\vx,\chi)\rvert_{\vx =\vx^{(0)}} \Delta \vx^{(1)}_i+ \mathcal{O}\[(\nabla \Psi)^3\].
\end{equation}
Inserting this expansion in Eq.~\ref{eq:vx}, we find the next to leading order term 
\begin{align}
\label{Deltax1}
\Delta \vx^{(1)}(\vtheta, \chi)& =  - 2\int_0^{\chi}\d \chi' \frac{\(\chi-\chi'\)}{\chi'} \nabla_{\vtheta} \Psi\(\vtheta\chi',\chi'\) = \chi \nabla_{\vtheta} \phi(\vtheta,\chi),
\end{align}
where we have identified the lensing potential $\phi$ (defined in Eq.~\ref{eq:lenspot}) in the last step.
At second order in the separation, we get
\begin{align}
\label{Deltax2}
\Delta \vx^{(2)}(\vtheta, \chi) & = - 2\int_0^{\chi}\d \chi' \(\chi-\chi'\) \left. \nabla_{\vx}\nabla_{\vx_j} \Psi\(\vx,\chi'\)\right\vert_{\vx=\vtheta\chi'}\, \Delta \vx^{(1)}_j(\vtheta, \chi')\\
& = - 2\int_0^{\chi}\d \chi' \frac{\chi-\chi'}{\chi'} \nabla_{\vtheta}\nabla_{\vtheta_j} \Psi\(\vtheta \chi',\chi'\) \nabla_{\vtheta_j} \phi (\vtheta, \chi').
\end{align}
We limit the discussion to terms second order in the lensing deflection here, but note that at third order we would have to take into account terms proprtional to $\(\Delta \vx^{(1)}\)^2$ and $\Delta \vx^{(2)}$. 

By comparison with Eq.~\ref{simple_lensing}, we can now identify the leading order contributions to the deflection angle $\Delta \vtheta$,
\begin{align}
\label{delta1}
\Delta \vtheta^{(1)}(\vtheta,\chi) & = - 2 \int_0^{\chi}\d \chi' \frac{\chi-\chi'}{\chi \chi'} \nabla_{\vtheta} \Psi\(\vtheta\chi',\chi'\) = \nabla_{\vtheta} \phi(\vtheta,\chi).
\end{align}
This is the well known Born approximation, which sums all deflections in the line-of-sight direction of the observer. In the Born approximation all deflections get projected onto a single lens plane.
At next to leading order, we add corrections from considering two consecutive lens planes. The image of the source after the first lens plane becomes the source for the next lens plane,
\begin{align}
\label{delta2}
\Delta\vtheta^{(2)}(\vtheta, \chi)  = - 2\int_0^{\chi}\d \chi' \frac{\chi-\chi'}{\chi \chi'} \nabla_{\vtheta}\nabla_{\vtheta_j} \Psi\(\vtheta \chi',\chi'\) \nabla_{\vtheta_j} \phi (\vtheta, \chi').
\end{align}
In the main text, we use expressions \ref{Deltax1}-\ref{delta2} as starting points for deriving higher order lensing corrections to observables such at the observed galaxy density or the lensing convergence. A more complete approach to expanding the lens equation can also be found in Ref.~\citep{2015JCAPFanizza}.
\subsection{Higher order lensing corrections to the projected galaxy field}
\label{app:gal_lens}

We relate the lensed galaxy overdensity field, $\tilde{\delta}_g$, to the unlensed overdensity field, $\delta_g$,  by
\beq
\label{delta_gal}
\tilde \delta_g\(\vtheta, z\) = \frac{\tilde n(\vtheta,z)-\bar{n}(z)}{\bar{n}(z)} =\frac{\left|\frac{\partial \vbeta}{\partial \vtheta}\right| n(\vbeta,z)-\bar{n}(z)}{\bar{n}(z)} = J(\beta, z)\[\delta_g(\vbeta,z)+1\]-1,
\eeq
with $n$ the galaxy number density (a tilde is used to denote lensed quantities), $\bar{n}$  its spatial mean and $J:=\left|\frac{\partial \vbeta}{\partial \vtheta}\right|$ the determinant of the lens-remapping, which is to be evaluated at position $\vbeta$ and at redshift $z$. Note that the area change $A=1/J$ introduced by lensing will result in $\tilde \delta_g\(\vtheta, z\)\neq 0$, even if the unlensed density field is homogeneous, $\delta_g(\vbeta,z)=0$.
The 2-dimensional projected lensed galaxy density $\tilde\delta_g(\vtheta)$ follows from Eq.~\ref{delta_gal} by  line-of-sight integration
\beq
\tilde\delta_g(\vtheta)=\int_0^{\chi_\max} \d \chi\,W_g(\chi) \tilde\delta_g\[\vtheta,z(\chi)\]
\eeq
Eq~\ref{delta_gal} assumes that the number of galaxies is conserved by lensing. However, in an actual survey with a magnitude limit $m_{\li}$, the (de-)focusing effect of lensing can bring galaxies (below) above the detection threshold. The combination of the two lensing effects - magnitude change and change of area - is well known as magnification bias ~\citep{1980Turner,1984Turner,1988Webster,1988Fugmann,1989Narayan,1989Schneider,1995Villumsen}.
It can be modeled as an effective change to the lensing Jacobian
\beq
J \rightarrow J^{1-2.5s}
\eeq
where $s$ is the change of number counts with magnitude at the magnitude limit $m_\li$. 
\begin{equation}
s= \left.\frac{\d \log_{10} n(m)}{\d m}\right|_{m_\li}.
\end{equation}
The slope $s$ is generally redshift dependent. Typical values for LSST range between $0.2$ and $0.4$ depending on the mean redshift of the sample. An additional size cut on the galaxies introduces a size bias which sources another effective increase in $s$.

Including magnification bias, the lensed galaxy density contrast is
\beq
\label{lensedgal_z}
\tilde{\delta}_g(\vtheta,z)=J^{1-2.5s}(\vbeta,z)\[\delta_g\(\vbeta,z\)+1\]-1=J^{1-2.5s}(\vtheta+\Delta\vtheta,z)\[\delta_g\(\vtheta+\Delta \vtheta,z\)+1\]-1.
\eeq
Projecting the above equation along the line-of-sight and rewriting in terms of co-moving coordinates, $\[\vx(\chi,\vtheta),\chi(z)\]$, gives
\beq
\label{lensedgal_proj}
g(\vtheta)=\int_0^{\chi_{\max}} \d\chi\, W_g(\chi) 
\left\{J^{1-2.5s} (\vx^{(0)}+\Delta\vx,\chi) \left[\delta_g\(\vx^{(0)}+\Delta\vx,\chi\)+1\right]-1\right\}.
\eeq
Note that we have chosen the same notation as in Appendix~\ref{app:lens_theory}, splitting the comoving separation $\vx$ in a no-lensing contribution $\vx^{(0)}=\vtheta \chi$ and a lensing correction $\Delta\vx$. We can now proceed in the same way as we did for the lensing convergence and expand in powers of $\Delta \vx$,
\begin{align}
\label{lensedgal_expa}
g(\vtheta) = \int_0^{\chi_s} \d \chi \,  W_g(\chi) \[J^{1-2.5s} \(1+ \delta_g + \delta_{g,i}  \Delta \vx_{i} + 
\frac{1}{2} \delta_{g,{ij}} \Delta \vx_{i} \Delta \vx_{j} \) -1 \]+\mathcal{O}(\Delta \vx_\perp^3)
\end{align}
We want to keep all terms up to second order in the lensing correction in the above equation. Assuming $\nabla^2 \Psi\propto\delta_m$ this will leave us with terms up to third order in the density contrast. 
Note that we have not yet expanded the determinant $J$ in Eq.\eqref{lensedgal_expa}. To do so, we write
\beq
J=\[(1-\kappa)^2-|\gamma|^2\]^{1-2.5s} \equiv 1+\Delta J=1+\Delta J^{(1)}+\Delta J^{(2)}+\Delta J^{(3)}+\mathcal{O}\(\delta_m\),
\eeq
and identify
\beq
\Delta J^{(1)}(\vtheta,\chi) = 5(s-0.4) \kappa(\vtheta, \chi).
\eeq
We further need to expand $\Delta \vx_\perp$, which sums the effect of many lensing events and is therefore non-linear in the lensing. As in Appendix~\ref{app:lens_theory} we write schematically (with expressions given in Eqs.~\ref{Deltax1} and~\ref{Deltax2}),
\beq
\Delta \vx=\Delta \vx^{(1)}+\Delta \vx^{(2)}+\mathcal{O}(\delta_m^3).
\eeq
Collecting all terms up to third order in the linear density field, results in
\begin{align}
\label{eq:gal_exp}
\nonumber
g(\vtheta)
=\, \int_0^{\chi_s} d\chi\, W_g(\chi) & \left[\delta_g(\vtheta\chi, \chi) + \delta_{g,i}(\vtheta\chi, \chi) \Delta \vx_i^{(1)}(\chi, \vtheta)  + \delta_{g,i}(\vtheta\chi, \chi) \Delta \vx_{i}^{(2)}(\chi, \vtheta)\right.  \\ 
\nonumber
& + \frac{1}{2} \delta_{g,{ij}}(\vtheta\chi, \chi) \Delta \vx_{i}^{(1)}(\chi, \vtheta)  \Delta \vx_{j}^{(1)}(\chi, \vtheta)  \\
\nonumber
& + \Delta J^{(1)} (\chi, \vtheta)\[1+\delta_g(\vtheta\chi, \chi) + \delta_{g,i}(\vtheta\chi, \chi) \Delta \vx_{i}^{(1)}(\chi, \vtheta) \]\\
\nonumber
& + \Delta J^{(2)}(\chi, \vtheta)\[1+\delta_g(\vtheta\chi, \chi)\]\\
& + \left.\Delta J^{(3)}(\chi, \vtheta) \right] +\mathcal{O}(\delta_m^4).
\end{align}
All of these terms must be taken into account when computing up to fourth order lensing corrections to the galaxy-galaxy and galaxy-lensing correlation functions. This is not only difficult because of the sheer number of terms (note that each $\Delta J^{(n)}$, with $n>1$ consists of several terms), but also because of the complicated structure of the  individual contributions.
In this work we are mainly interested in the question whether we can recover similar cancellation between higher order terms in the cross correlation as in lensing auto correlation and leave the estimation of additional terms for future work.
\section{Appendix: Expressions for some additional Jacobian terms}
\label{app:jacobian_terms}
In this appendix we give expressions for some of the correction terms that involve the lensing Jacobian.
Correlating $J^{(1)}\delta^{(2)}$ with $\kappa^{(1)}$ gives
\begin{align}
C_{1 2 \rm{J}1}^{(\kappa g)}(L) = \frac{ 5(s-0.4)}{4} \int \frac{\d^2 \vl}{(2 \pi)^2} l^2 L^4 \int_0^{\chi_\max} \d \chi \, W_g(\chi) C_L^{\phi \phi}(\chi,\chi_\CMB) C^{\phi\delta_h}_{l}(\chi).
\end{align}
Similarly for the correlation between $J^{(1)} \delta^{(1)} $ and $\kappa^{(2)}$
\begin{align}
C_{2 1 \rm{J}1}^{(\kappa g)}(L) & =-\frac{5(s-0.4)}{2} \int \frac{\d^2 \vl}{(2 \pi)^2} l^2 \(\VL\cdot \vl\) \[\vl\cdot\(\VL- \vl\)\] \\ 
\nonumber 
& \int_0^{\chi_\max} \d \chi \, W_g(\chi) \int_0^{\chi_\CMB} \d \chi'' W_\kappa(\chi'', \chi_\CMB) C_{|\VL-\vl|}^{\phi\Psi}(\chi,\chi'') C_l^{\delta\phi}(\chi,\chi'')\\
& +\frac{5(s-0.4)}{2} \int \frac{\d^2 \vl}{(2 \pi)^2} l^2  \[\VL\cdot\(\VL- \vl\)\]\[\vl\cdot\(\VL- \vl\)\] \\
\nonumber
& \int_0^{\chi_\max} \d \chi \, W_g(\chi)\int_0^{\chi_\CMB} \d \chi'' W_\kappa(\chi'', \chi_\CMB) C_l^{\phi\phi}(\chi,\chi'') C_{|\VL-\vl|}^{\delta\Psi}(\chi,\chi'').
\end{align}
The first order lensing correction to the Jacobian,$J^{1,1}$ can be contracted with  $\delta^1$ and $\kappa^{1}$. The corresponding term is
\begin{align}
\nonumber
    C_{1 1 \rm{J}11}^{(\kappa g)}(L) = & 5 (s-0.4)\, L^2 \int \frac{\d^2 \vl}{(2 \pi)^2} \[(\vl+\VL)\cdot \VL\] (\VL \cdot \vl) \int_0^{\chi_\max} \d \chi\, W_g(\chi) \\
    & \int_0^{\chi} \d \chi' W_\kappa(\chi', \chi) \int_0^{\chi_\CMB} \d \chi'' W_\kappa(\chi'', \chi_\CMB) C_l^{\Psi\Psi}(\chi',\chi'') C_L^{\phi\delta}(\chi',\chi)\\
    \nonumber
    &  5 (s-0.4)\, L^2 \int \frac{\d^2 \vl}{(2 \pi)^2} \[(\vl+\VL)\cdot \vl\] l^2  \int_0^{\chi_\max} \d \chi\, W_g(\chi) \\
    & \int_0^{\chi} \d \chi' W_\kappa(\chi', \chi) \int_0^{\chi_\CMB} \d \chi'' W_\kappa(\chi'', \chi_\CMB) C_l^{\phi\Psi}(\chi',\chi'') C_L^{\delta\Psi}(\chi,\chi'')
\end{align}
At second order in the Jacobian we have further the correlation of $\kappa^2 \delta^1$ with $\kappa^1$
\begin{align}
\nonumber
    C_{1 1 \rm{J}2a}^{(\kappa g)}(L) = & \frac{ 5 (s-0.4)}{2} L^2 \int \frac{\d^2 \vl}{(2 \pi)^2} l^4 \int_0^{\chi_\max} \d \chi\, W_g(\chi) \int_0^{\chi} \d \chi' W_\kappa(\chi', \chi) \int_0^{\chi} \d \chi''\, W_\kappa(\chi'', \chi) \\
    &  \int_0^{\chi_\CMB} \d \chi'''\, W(\chi''', \chi_\CMB)  C_l^{\Psi\Psi}(\chi',\chi'') C_L^{\delta\Psi}(\chi,\chi''')\\
    \nonumber
    & \frac{ 5 (s-0.4)}{2} L^4 \int \frac{\d^2 \vl}{(2 \pi)^2} l^2 \int_0^{\chi_\max} \d \chi\, W_g(\chi) \int_0^{\chi} \d \chi' W_\kappa(\chi', \chi) \int_0^{\chi} \d \chi''\, W_\kappa(\chi'', \chi) \\ 
    &  \int_0^{\chi_\CMB} \d \chi'''\, W(\chi''', \chi_\CMB) \[C_l^{\Psi\delta}(\chi',\chi) C_L^{\Psi\Psi}(\chi'',\chi''')+ C_l^{\Psi\delta}(\chi'',\chi) C_L^{\Psi\Psi}(\chi',\chi''')\].
\end{align}
The two other $J^{(2)}$ terms, i.e. $\gamma_1^2 \delta^1 \kappa^1$ and $\gamma_2^2 \delta^1 \kappa^1$, only differ to this in their derivative structure.
For $\gamma_1^2 \delta^1 \kappa^1$, we get in the first line (assuming that we can align the $x$-axis with the exterior $\VL$ vector, and that $\varphi$ is the angle between $\vl$ and this $x$-axis, 
\begin{equation}
    -\frac{1}{4} L^2 \[l^2(\sin^2\varphi-\cos^2\varphi)\]^2
\end{equation}
and in the second line
\begin{equation}
    -\frac{1}{4} L^4 \[l^2(\sin^2\varphi-\cos^2\varphi)\].
\end{equation}
For $\gamma_1^2 \delta^1 \kappa^1$, only the first term has non-odd parity and is non-zero,
\begin{equation}
    -L^2 l^4 \sin^2\varphi \cos^2\varphi.
\end{equation}
\section{Appendix: Odd parity terms in Section~\ref{sec:limber}}
\label{app:add_terms_limber}
The following terms integrate to zero, because the integrand in the integral over $\VL_1$ changes sign under $\VL_1\leftrightarrow-\VL_1$
\begin{align}
\nonumber
C_{31A}^{(\kappa g)}(L)=&4 \ \mathcal{A}^3 \int_{\VL_1} \frac{\[\VL\cdot \VL_1\]^3}{L_1^4 L^2} \int_0^{\chi_\max} \drm \chi\, \frac{W_\kappa\[\chi,p(\chi_s)\]W_g(\chi)}{\chi^2}\, P_{mg}(L/\chi,\chi) \\
\nonumber
& \times  \int_0^\chi \drm \chi'\,  \frac{\[W_\kappa(\chi',\chi)\]^2}{\chi'^2} \, P_{mm}(L_1/\chi',\chi')\\
=&0\,  (\text{odd parity})
\end{align}
\begin{align}
\nonumber
C^{\kappa g}_{13c}(L') = & - 4 \mathcal{A}^3 \int \frac{\d^2 \VL_1}{(2\pi)^2} \frac{\VL_1\cdot \VL'}{L_1^2} \int_0^{\chi_s} \d \chi \frac{W_g(\chi) W_\kappa\(\chi,\chi_s\)}{\chi^2} \\
& \int_0^{\chi} \d \chi' \frac{W_\kappa(\chi',\chi) W_\kappa(\chi',\chi_\gal)}{\chi'^2} P_{mg}(L'/\chi,\chi)  P_{mm}(L_1/\chi',\chi')\\
\nonumber
& +4 \mathcal{A}^3 \int \frac{\d^2 \VL_1}{(2\pi)^2} \frac{\VL_1\cdot \VL'}{L_1^2} \int_0^{\chi_s} \d \chi \frac{W_g(\chi) W_\kappa(\chi',\chi_\gal)}{\chi^2} \\
& \int_0^{\chi} \d \chi' \frac{W_\kappa\(\chi,\chi_s\) W_\kappa(\chi',\chi) }{\chi'^2} P_{mg}(L'/\chi,\chi)  P_{mm}(L_1/\chi',\chi')\\
 = & \,  0 \mathrm{\, (parity)}.
\end{align}
\bibliographystyle{plain}
\bibliography{CMBLens,Lens,Zotero,LSS}
\end{document}